\documentclass[aps,prl,twocolumn,floatfix,superscriptaddress]{revtex4-1}

\usepackage{rotating}
\usepackage{epstopdf}
\usepackage{graphicx}
\usepackage{natbib}
\usepackage{epstopdf}
\usepackage{amsmath}
\usepackage{amssymb}
\usepackage{mathbbol}
\usepackage{xcolor}
\usepackage{lipsum}
\usepackage{amssymb}
\usepackage{mwe}
\usepackage{lipsum} 
\usepackage{siunitx}
\usepackage[normalem]{ulem}
\usepackage{float}

\usepackage{empheq}
\usepackage{afterpage}
\usepackage[colorlinks,citecolor=blue,urlcolor=red]{hyperref}
\usepackage{cleveref}

\newcommand{\beq}{\begin{equation}}
\newcommand{\eeq}{\end{equation} \smallskip}
\newcommand{\beqy}{\begin{eqnarray}}
\newcommand{\eeqy}{\end{eqnarray} \smallskip}

\newcommand{\bit}{\begin{itemize}}
\newcommand{\eit}{\end{itemize}}
\newcommand{\bmat}{\begin{pmatrix}}
\newcommand{\emat}{\end{pmatrix}}
\newcommand{\red}[1]{\textcolor{red}{#1}}

\begin{document}
	
	\title{Kibble-Zurek  mechanism in driven-dissipative systems crossing a non-equilibrium phase transition}
	
	\author{A. Zamora}
	\thanks{These three authors contributed equally to this work.}
	\address{Department of Physics and Astronomy, University College London,
		Gower Street, London, WC1E 6BT, United Kingdom}
	
	\author{G. Dagvadorj} 
	\thanks{These three authors contributed equally to this work.}
	\address{Department of Physics and Astronomy, University College London,
		Gower Street, London, WC1E 6BT, United Kingdom}
	\address{Department of Physics, University of Warwick, Coventry, CV4 7AL, United Kingdom}
	
	\author{P. Comaron}
	\thanks{These three authors contributed equally to this work.}
	\address{Joint Quantum Centre (JQC) Durham-Newcastle, School of Mathematics, Statistics and Physics, Newcastle University, Newcastle upon Tyne, NE1 7RU, United Kingdom}
	\address{Institute of Physics, Polish Academy of Sciences, Al. Lotników 32/46, 02-668 Warsaw, Poland}
	
	\author{I. Carusotto}
	\address{INO-CNR BEC Center and Universit\`a di Trento, via Sommarive 14, I-38123 Povo, Italy}
	
	\author{N. P. Proukakis}
	\address{Joint Quantum Centre (JQC) Durham-Newcastle, School of Mathematics, Statistics and Physics, Newcastle University, Newcastle upon Tyne, NE1 7RU, United Kingdom}
	
	\author{M. H. Szyma\'nska}
	\address{Department of Physics and Astronomy, University College London,
		Gower Street, London, WC1E 6BT, United Kingdom}

	\begin{abstract}
		The Kibble-Zurek mechanism constitutes one of the most fascinating and universal phenomena in the physics of critical systems. It describes the formation of domains and the spontaneous nucleation of topological defects when a system is driven across a phase transition exhibiting spontaneous symmetry breaking. While a characteristic dependence of the defect density on the speed at which the transition is crossed was observed in a vast range of equilibrium condensed matter systems, its extension to intrinsically driven-dissipative systems is a matter of ongoing research.
		In this work we numerically confirm the Kibble-Zurek mechanism in a paradigmatic family of driven-dissipative quantum systems, namely exciton-polaritons in microcavities. Our findings show how the concepts of universality and critical dynamics extend to driven-dissipative systems that do not conserve energy or particle number nor satisfy a detailed balance condition.
	\end{abstract}

	\maketitle
	
	One of the most intriguing universal phenomena encountered in  {the physics of critical systems} is the so-called Kibble-Zurek (KZ) mechanism, which successfully describes the  {spontaneous} appearance of long-lived topological defects in complex systems that undergo a spontaneous symmetry breaking when crossing a critical point  {at a finite speed}~\cite{kibble1976topology,zurek1985cosmological}.
	This mechanism is general and spans across vastly different physical realisations and length/energy scales, with topological defects ranging from monopoles and vortices to strings and domain walls, depending on the symmetries and the spatial dimensions.  In spite of this variety, the density of topological defects  {has a universal dependence on} the rate of change of the  {control parameter across the transition} and  {on}  {the} critical exponents of the system~\cite{kibble1976topology,zurek1985cosmological,del2013causality,dziarmaga2010dynamics,biroli2010}. 
	
	This phenomenon can be  physically understood by considering the different stages of critical dynamics when the control parameter is scanned across the critical point. 
	In the initial stages of the dynamics, far from the critical point, the system exhibit an {\em adiabatic} behaviour permitted by the fact that the characteristic relaxation time $\tau$  is much shorter than the characteristic time of the control parameter ramp.
	{Later on, since the characteristic relaxation time $\tau$ diverges at the critical point, there must necessarily exist a time after which the system is no longer able to readjust itself adiabatically following the variation of the control parameter and thus enters into a so-called \emph{impulse} regime. According to the KZ picture, the density of the topological defects that are left behind at this point of the evolution is determined by the correlation length of the system at this \emph{`crossover time'}~\cite{zurek1985cosmological}.
		
		{This} KZ mechanism, first proposed in the cosmological context~\cite{kibble1976topology,zurek1985cosmological}, has been  studied in vastly different contexts spanning across superconducting junction arrays, ion crystals, quantum Ising chains, classical spin systems,  {holographic superconductors}, fermionic and bosonic atomic and helium superfluids, and cosmological scenarios~\cite{yates1998vortex,laguna1997density,dziarmaga2008winding,kolodrubetz2012nonequilibrium,jelic2011quench,sonner2015universal,chesler2015defect,morigi-KZ-2016,zurek-soliton-2010,liu-KZ-2018,dora-KZ-2019}, with direct experimental confirmations in a broad range of different physical systems~\cite{chuang1991cosmology,bowick-liquidcrystals-1994,hendry-he-4,bauerle1996laboratory,ruutu1996vortex,maniv2003observation,pyka2013topological,deutschlander-colloids-2015,stamper-kurn-2006,weiler2008spontaneous,lamporesi2013spontaneous,dalibard-ring-KZ-2014,dalibard-2D-KZ-2015,hadzibabic-KZ-2015,shin-KZ-Fermi-2019}. 
		A common feature of these studies is that they mostly address cases that are at, or close to, thermal equilibrium and conserve energy and particle number.

		Recent experimental progress in the study of exciton-polaritons in semiconductor microcavities embedding quantum wells~\cite{carusotto2013quantum,deng2010exciton,kasprzak2006bose} -- henceforth referred to as polaritons -- has led to hybrid light-matter systems which exhibit a condensation phase transition and the spontaneous appearance of a macroscopic coherence while being inherently in a strongly non-equilibrium condition~\cite{carusotto2013quantum}, as the system requires an external pump to compensate for the losses by continuously injecting new polaritons. 
		Polaritons therefore constitute excellent physical platforms to explore the influence that the non-conservation of energy and particle number, and the breaking of the detailed balance condition, may have on the critical dynamics.
		Pioneering works have started addressing the new features exhibited by the ordered state~\cite{szymanska2006nonequilibrium,PhysRevLett.99.140402}, by the non-equilibrium phase transition~\cite{sieberer2016keldysh,sieberer2015thermodynamic,PhysRevX.5.011017,zamora2017tuning}, the extension of the adiabaticity concept to non-equilibrium scenarios~\cite{hedvall2017dynamics,tomka2018accuracy,Verstraelen2020}, the spontaneous formation of defects 
		under a time-dependent pump~\cite{lagoudakis2011probing,matuszewski2014universality,solnyshkov2016kibble}, and the late-time relaxation past a sudden quench~\cite{PhysRevB.95.075306,comaron2018dynamical}.

		In this Letter we investigate the KZ mechanism in the non-equilibrium phase transition, focusing, in contrast to previous studies \cite{lagoudakis2011probing,matuszewski2014universality,solnyshkov2016kibble,PhysRevB.95.075306,comaron2018dynamical}, on the characteristic dependence of the spontaneous vortex nucleation process on the switch-on rate of the pump.
		Our numerical results provide a direct evidence of the adiabatic-to-impulse crossover and confirm the validity of the KZ picture also in the driven-dissipative context of a non-equilibrium phase transition.  
		Compared to a direct study of the number of vortices that are still present at the end of the ramp as a function of the ramp speed, our approach has the key advantage of being insensitive to those vortex annihilation processes that may occur past the critical point~\cite{biroli2010} and were shown to contribute to the late-time phase ordering dynamics studied in~\cite{comaron2018dynamical}.
		
		The key idea for testing and demonstrating the KZ mechanism is to numerically simulate the dynamical evolution and extract from it the `crossover time' (subsequently referred to as $-\hat{t}$)  {after which the system is no longer able to adiabatically follow the steady-state corresponding to the instantaneous value of the pump. This value is then compared} to the  {corresponding} prediction  {of the KZ model,  {i.e.~to} the time at which the speed of variation of the control parameter starts exceeding the characteristic relaxation time of the system. Similar strategies were previously used for equilibrium scenarios in~\cite{yates1998vortex,laguna1997density,dziarmaga2008winding,jelic2011quench}.}
		
		In order to validate the universality of the critical polariton dynamics, we  perform two independent calculations for the two most celebrated pumping schemes,
		which differ in their method of injection and subsequent relaxation processes leading to condensation \cite{carusotto2013quantum}; specifically, we consider the optical parametric oscillation (OPO) scheme, and  the incoherent pumping (IP) scheme (see Ref.~\cite{supp_mat} for details).
		
		\paragraph*{Polariton Phase Transition \& Modelling.}
		{As discussed in the literature on spontaneous macroscopic coherence and the non-equilibrium condensation phase transition of polaritons}~\cite{carusotto2013quantum,carusotto2005spontaneous,PhysRevB.93.195306,comaron2018dynamical,dagvadorj2015nonequilibrium,dunnett2018properties,chiocchetta2013non,caputo2018topological}, both the OPO and the IP polariton systems show  {rich yet qualitatively very similar phase diagrams}, with two  {main distinct phases}: i) a disordered phase  {displaying a low} density of polaritons,  {an} exponential decay of  {spatial} correlations and a plasma of unbound vortices; ii) a (quasi)ordered phase  {displaying a significant} density of polaritons,  {an algebraic decay of spatial correlations (at least up to relatively long distances~\cite{dagvadorj2015nonequilibrium,PhysRevX.5.011017,keeling2017superfluidity}}) and a low density of vortices,  {mostly bound in vortex-antivortex pairs}~\cite{comaron2018dynamical,dagvadorj2015nonequilibrium,caputo2018topological}.  
		
		\begin{figure}
			\includegraphics[width=\linewidth]{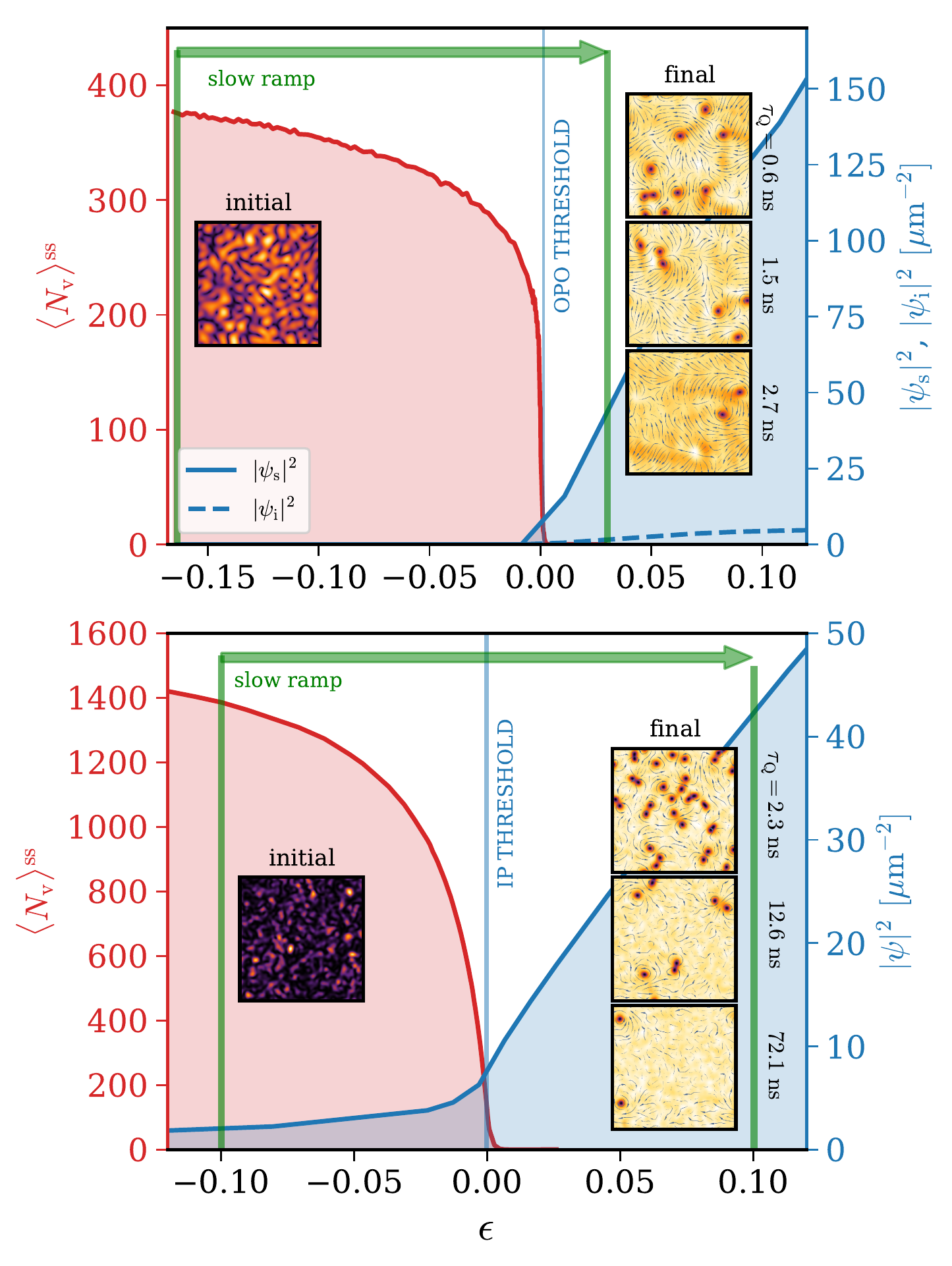}
			\caption{{{ {Non-equilibrium} phase transition in the polariton  {system}}.}
				Top panel:  {OPO case. Steady-state,} noise-averaged densities for the signal $|\psi_\textrm{s}|^2$ (blue solid line) and idler $|\psi_\textrm{i}|^2$ (blue dashed line) fields.
				Bottom panel:  {IP case. Steady-state,} noise-averaged field density $|\psi|^2$ (blue solid line). 
				{In both panels the red curves indicate the steady-state, noise-averaged} number of topological defects $\langle N_\mathrm{v} \rangle^{\mathrm{ss}}$.  {All quantities are plotted as a function of the distance to criticality $\epsilon$}.
				The  {insets} show typical snapshots of the  {field profile in the initial and final states, as indicated by the thick green arrows at the top of the panels. Typical final state profiles are displayed for different ramp speeds of different timescale $\tau_Q$.} 	}
			\label{fig:IP_OPO_spectrum}
		\end{figure}

		{The intensity of the pump, namely $f_p$ (for OPO) and $P$ (for IP), acts as a control} parameter  {and} the system  {is} driven from one phase to the other  {by simply ramping up its value in time at different rates. As usual in condensation phase transitions, the transition from the disordered to the (quasi)ordered phase is accompanied by the breaking of the U(1) symmetry associated with the phase of the polariton condensate. In the present 2D case, it can be pictorially understood as being mediated by the unbinding of vortex-anti-vortex pairs into a plasma of free vortices~\cite{minnhagen1987two,dagvadorj2015nonequilibrium,PhysRevX.5.011017,wachtel2016electrodynamic}.}
		A schematic of the phase transition process, depicting our quench sequence and typical initial and final snapshots of the polariton field are shown in Fig.~\ref{fig:IP_OPO_spectrum}.
		
		\begin{figure}
			\includegraphics[width=\linewidth]{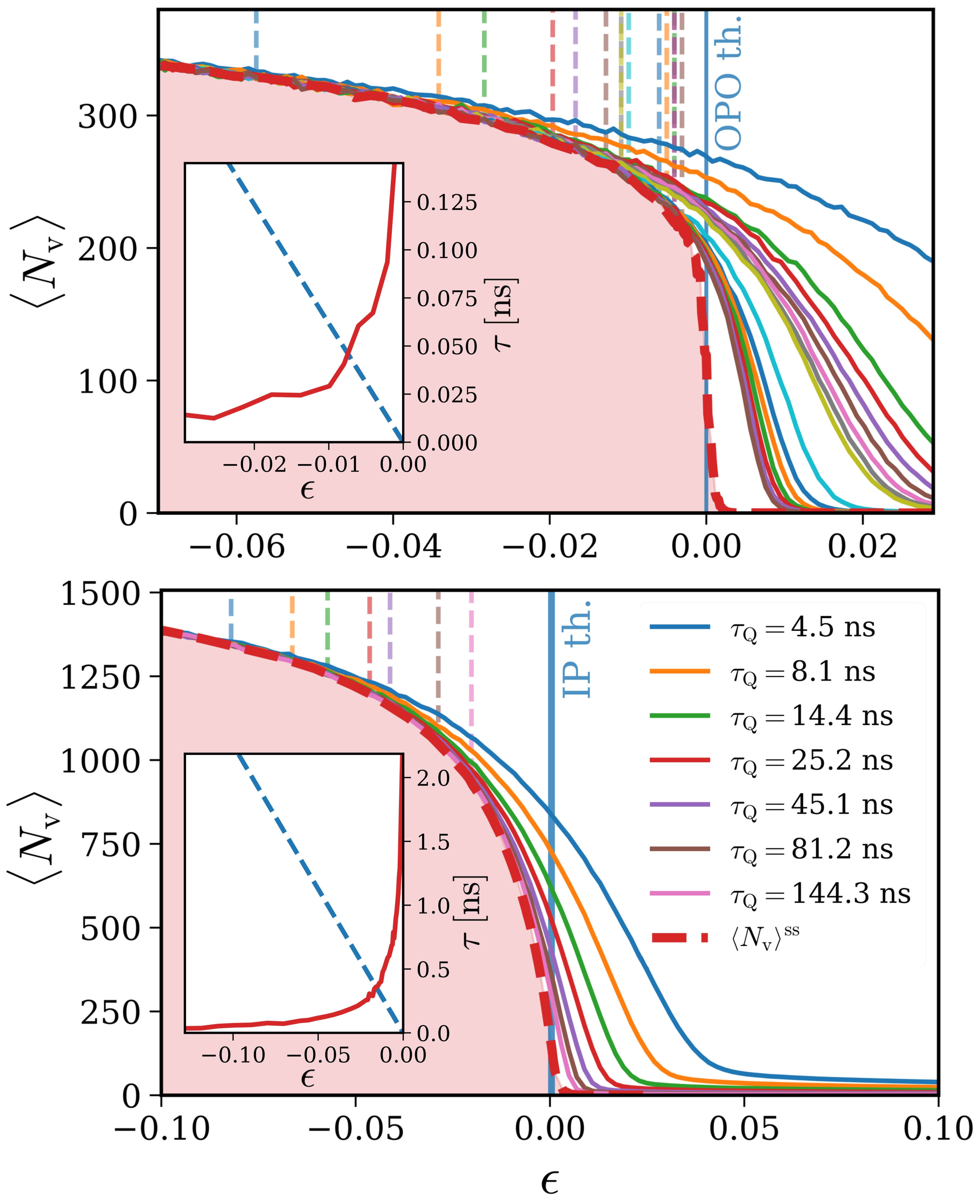}
			\caption {  
				Average number of vortices as a function of the distance to criticality $\epsilon$ for different ramp speeds of characteristic time $\tau_\mathrm{Q}$ (thin  {solid} curves), and at steady-state  {(red dashed curves)} for OPO (top) and IP (bottom panels) pumping schemes.
				Chosen values for OPO {(from right to left)}: $\tau_\textrm{Q} = 0.3$ (blue curve),  $0.6$,   $0.9$,   $1.2$,   $1.5$,   $1.8$,   $2.1$,   $2.4$,   $2.7$,   $3$,   $6$,    $9$,	$12$,   $15$,   $18$,   $21$,   $24$ $\mathrm{ns}$  {(brown curve)}
				(with corresponding IP values indicated within figure).
				{For each value of $\tau_Q$,} the dashed vertical lines  {indicates the point} $\hat{\epsilon}_\textrm{num}(\tau_\mathrm{Q})$  {where the number of vortices in the dynamical evolution starts departing} from the steady-state value (see Ref.~\cite{supp_mat} for details).
				{\textbf{Insets:} red solid curves show the characteristic relaxation time, $\tau$, of the vortices  as a function of the distance to criticality, $\epsilon$.
					Dashed straight line indicates an example of the dependence of the characteristic time  $\epsilon(t)/ \dot{\epsilon}(t)$ on $\epsilon$ for a specific choice of ramp parameters, namely $\tau_Q=
					1.2~$ns (OPO) and $\tau_Q =4.5~$ns (IP) with $a_p=0.1942$ (OPO) and $a_p=0.2$ (IP). 
				}}
				\label{fig:vortices_vs_epsilon}
			\end{figure}
			
			{A powerful way to theoretically describe the collective dynamics of the polariton field across the phase transition is based on} a generalized  {stochastic} Gross-Pitaevskii  equation.  {In this model,} the nonlinearity arises from the effective  {polariton-polariton} interactions,  {with suitable additional terms included to describe pumping and losses, and stochastic noise which accounts for} the quantum fluctuations~\cite{carusotto2005spontaneous,PhysRevLett.99.140402,PhysRevB.79.165302,carusotto2013quantum}. 
			A detailed description of such equations for the polaritons can be found in Ref.~\cite{supp_mat}.  {In order to focus on the intrinsic features of the KZ physics, we restrict our investigation here to the simplest case of a spatially homogeneous system with periodic boundary conditions. }
			
		}
		\paragraph*{Ramp Protocol.}
		{For the OPO (IP) case}, we drive the polariton system through  {the non-equilibrium phase transition by ramping in time the pump intensity {$f_p(t)$ [$P(t)$]} across the critical value {$f_p^c$ ($P_c$)}, starting from an initial steady-state at a pump intensity {$f_p^i$ ($P_i$)} in the disordered phase to a pump intensity {$f_p^f$ ($P_f$)} well in the (quasi-)ordered phase. 
			The ramp follows a linear law of characteristic time $\tau_\mathrm{Q}$. 
			We characterise the phase transition in terms of} the distance to criticality, which is  {quantified by the time-dependent parameter $\epsilon(t)$ defined as}
		{\begin{equation}
			\epsilon(t)= \left \{
			\begin{aligned}
			& \frac{f_{p}^{c}-f_p(t)}{f_{p}^{c}}, \ \ \textrm{(for OPO)} \\
			& \frac{P(t)-P_c}{P_c} , \qquad  \text{(for IP)}
			\end{aligned} \right\}
			=\varsigma \left( \frac{a_p}{\tau_Q} \right)
			t 
			\label{eq:finite_quench}
			\end{equation}}
		where the $a_p \equiv
		(f_{p}^{i}-f_{p}^{f}) / f_{p}^{c}$ (OPO) or $a_p \equiv (P_f-P_i) / P_c$ (IP) parameter is chosen to have the same value for all ramps within a given pumping scheme and the sign of $\varsigma=\mp 1$ (for OPO/IP) is chosen for
		consistency with the usual definition of the control parameter in the previous literature on phase transitions.
		Note that this distinction is required because for the OPO transition we are considering the upper threshold \cite{upper_th}, so we need to quench from high to low values of the
		control parameter.
		For convenience, we define the  {origin, $t=0$, of the time axis, as the time} when the system crosses the critical point, i.e. $\epsilon(t=0)=0$ based upon {$f_p(t=0)=f_p^c$ (OPO) and $P(t=0)=P_c$ (IP)}. Therefore, in both IP/OPO cases, the initial time of the simulation has a negative value, i.e. $t_i<0$.
		Further {details of the finite speed ramp} adopted can be found in Ref.~\cite{supp_mat}.
		
		\paragraph*{Testing the KZ mechanism.}
		
		First, we need to numerically determine the crossover time, $\hat{t}_{\rm num}$, from the vortex dynamics  during a finite-speed ramp.
		The number of vortices across the  {Berezinskii--Kosterlitz--Thouless} (BKT) transition at steady-state is known to decrease gradually as the transition is approached from the disordered side, and  {to} exhibit a sharp decrease in a narrow region around the critical point, as already analysed for OPO polaritons in~\cite{dagvadorj2015nonequilibrium,comaron2018dynamical,dunnett2018properties}.  {This feature, in combination with a simultaneous study of the  {spatial} correlation function is used to precisely locate the critical point.} This behaviour is shown for both OPO and IP cases in terms of the distance to criticality 
		$\epsilon$ by the dashed red lines in Fig.~\ref{fig:vortices_vs_epsilon} (with $\epsilon=0$  denoted by vertical solid lines).  
		When  {ramping the pump intensity} from the disordered phase, the vortex density  {initially} follows the steady-state density during the initial stages of the dynamics, $\epsilon \ll 0$. However, as the dynamical system cannot follow the steady-state through the  {critical point}, where the relaxation time diverges,  the vortex density  {eventually} \emph{departs} from its steady-state value,  {as shown by the solid lines for different ramp timescales, $\tau_\mathrm{Q}$.} From this plot  {we directly} extract the numerical crossover time, $\hat{t}_\textrm{num}<0$, at which  {each of the dynamical curves}  {starts to deviate} from the steady-state one. Such times  {are} highlighted for  {each value of} $\tau_\mathrm{Q}$ by  {a} vertical dashed  {line  {in Fig.~\ref{fig:vortices_vs_epsilon}}. These lines clearly}  {demonstrate a significant increase in the deviation for smaller values  of $\tau_\mathrm{Q}$, i.e.~for {\em faster} ramps.}
		More details of the extraction of $\hat{t}_\textrm{num}$ from the data and the dependence of $\hat{t}_\textrm{num}$ on $\tau_\mathrm{Q}$ can be found in Ref.~\cite{supp_mat}.
		
		In order to  {explicitly} verify the KZ mechanism,  {we should now compare the above numerical prediction for $\hat{t}_{\rm num}$ with the one} extracted by the KZ hypothesis, denoted here by $\hat{t}_{\rm KZ}$. 
		The KZ hypothesis states that  {the dynamical results should start departing from the corresponding steady-state ones} at the time, $\hat{t}_{\rm KZ}$,  {at which the relaxation time $\tau$ equals the timescale of the pump variation. Expressed in terms of the distance to criticality, $\epsilon(t)$:
			\begin{equation}
			\tau(\epsilon(t))_{t=\hat{t}_\textrm{KZ}} = A \left|\frac{\epsilon(t)}{\dot{\epsilon}(t)} \right|_{t=\hat{t}_\textrm{KZ}}.
			\label{eq:zurek}
			\end{equation}
			{Here,} the dependence of the crossover time $\hat{t}_\textrm{KZ}$ on the  {ramp speed} is contained in the time derivative of the distance to criticality $\epsilon(t)$,  {and $A$ is a constant parameter of order one.}
			The relaxation time $\tau$ is known to diverge at the critical point, and so the interesection of this with {the straight line} $\epsilon(t) / \dot{\epsilon}(t)$ defines the  {crossover time at which} the system crosses from an adiabatic to an impulse behaviour. 
			This is schematically represented in {the insets of Fig.~\ref{fig:vortices_vs_epsilon}}.  
			Changing the rate at which the pump intensity is varied will directly affect, {via \eqref{eq:finite_quench}}, the ramp speed, and thus set a different slope for $\epsilon(t) / \dot{\epsilon}(t)$.
			Dashed straight lines in the insets of Fig.~\ref{fig:vortices_vs_epsilon} depict an example of the dependence of the characteristic time  $\epsilon(t)/ \dot{\epsilon}(t)$ on $\epsilon$ (see caption of Fig.~\ref{fig:vortices_vs_epsilon} for the exact choice of parameters).
			Applying this protocol to different values of $\tau_Q$ gives the KZ prediction for $\hat{\epsilon}=\epsilon(\hat{t})$, based on Zurek's expression \eqref{eq:zurek} (with $A=1$).
			In turn, this defines a different intersection point with  the relaxation time, $\tau(\epsilon)$, in the $\tau(\epsilon)$ vs. $\epsilon$ plot {(insets of Fig.~\ref{fig:vortices_vs_epsilon})}.
			In order to extract the interesection points for different  {ramp speeds}, we thus need to first extract the system relaxation time, $\tau(\epsilon)$, plotted by the solid red line in {insets of Fig.~\ref{fig:vortices_vs_epsilon}}.  {For each value of $\epsilon$ in the disordered phase,} this is obtained by considering the relaxation time of the number of vortices $N_\mathrm{v}$  {to} the steady-state value $N_\mathrm{v}^\mathrm{ss}$  {after an infinitely rapid quench of $\epsilon$ towards the desired value},
			{\begin{equation}
				N_\mathrm{v}(t)-N_\mathrm{v}^\mathrm{ss} \propto \exp(-t/\tau(\epsilon)),
				\end{equation}}
			(see Ref.~\cite{supp_mat} for more details).
			
			\paragraph*{Validation of KZ mechanism for driven-dissipative systems.}
			
			The above procedure indicates a linear relation between the \emph{numerical} ( $\hat{t}_\textrm{num}$) and the \emph{predicted} ( $\hat{t}_\textrm{KZ}$) time for the crossover from adiabatic to impulse behaviour, as shown in Fig.~\ref{fig:t_num_vs_t_pred}.
			We have checked that such a linear relation holds for different choices of the proportionality constant $A$ (beyond $A=1$), thus confirming the independence of our conclusions on its specific choice (see Ref.~\cite{supp_mat} for more details).
			Since the KZ mechanism is based on the critical
			properties around the phase transition point, one can
			naturally expect it to be restricted to sufficiently slow ramps,  {for which the linear relation is clearly defined.} 
			On the other hand, significant deviations from
			the linear relation between  $\hat{t}_\textrm{num}$ and $\hat{t}_\textrm{KZ}$ are expected to arise for small values of $\hat{t}$, where non-universal corrections become important. A hint at such deviations is visible in the presented IP results. 
			{Note that}  {the (non-universal) {\em intercept} of the IP polariton system is highly sensitive to the {\em exact} location} of the critical point, since a tiny shift in the identification of the critical point  {within the critical region} can shift the intercept towards, or away from, a zero value.
			
			\begin{figure}[t]
				\includegraphics[width=\linewidth]{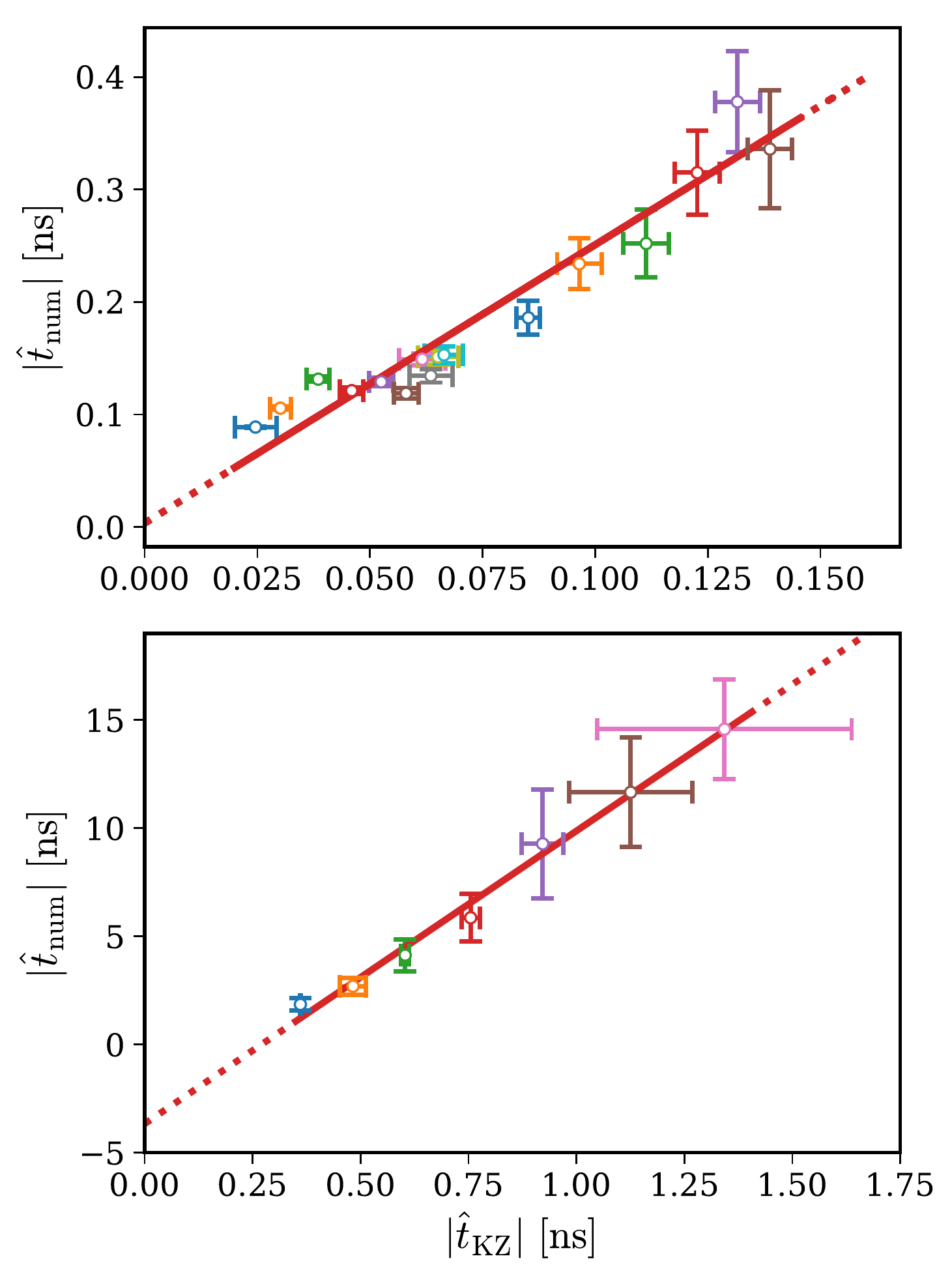}
				\caption {
					Numerical prediction for the crossover time $|\hat{t}_\textrm{num}|$ (corresponding to $\hat{\epsilon}_\mathrm{num}$ in Fig.~\ref{fig:vortices_vs_epsilon}) {plotted} with errorbars \cite{error_FIG4} as a function of the theoretical crossover time $|\hat{t}_\textrm{KZ}|$ {(see insets of  Fig.~\ref{fig:vortices_vs_epsilon})} predicted by the Zurek relation Eq.~\eqref{eq:zurek} (with $A=1$)
					for the OPO (top panel) and IP (bottom panel) pumping schemes.
					The observed linear dependence between both variables is a clear indication of the applicability of Zurek's relation Eq.~\eqref{eq:zurek}, and of the KZ mechanism. 
					We obtain a zero intercept (within the error bars) for the OPO polariton system and a small non-zero intercept for the IP case,  {which indicates non-universal sub-leading order corrections}.
				}
				\label{fig:t_num_vs_t_pred}
			\end{figure}

			\paragraph*{Conclusions.}
			
			{We have investigated the open question of} the extension of the Kibble-Zurek phenomenon to driven-dissipative quantum systems. Specifically, we  {have considered the dynamics of the vortex density during a spontaneous symmetry breaking process across a critical point for} a paradigmatic case of a non-equilibrium  {phase transition}, namely the  {condensation of exciton-polaritons in semiconductor microcavities embedding quantum wells in the strong light-matter coupling regime}.
			Our numerical findings, based on very accurate simulations of the dynamical equations of the systems for experimentally relevant parameters,  
			fully confirm the existence of a crossover from an adiabatic to an impulse behaviour at a point that depends on the ramp speed, and the validity of  Zurek's relation
			[Eq.~\eqref{eq:zurek}]. Our analysis thus shows that the KZ mechanism can maintain its validity even in the case of non-equilibrium phase transitions.
			
			\paragraph*{Acknowledgements. -}
			We would like to thank M. Matuszewski for fruitful discussions and A. Ferrier for a revision of the text.  We acknowledge financial support from the EPSRC (Grants No. EP/I028900/2 and No. EP/ K003623/2), and the Quantera ERA-NET cofund NAQUAS and InterPol projects (EPSRC Grant No. EP/R043434/1 and EP/R04399X/1).
			I.C. acknowledges financial support from the Provincia Autonoma di Trento and from the European Union via the FET-Open Grant ``MIR-BOSE'' (737017) and the Quantum Flagship Grant ``PhoQuS'' (820392).
			The data that support the findings of this work are available  {by following the link at https://doi.org/10.25405/data.ncl.10029515.v1}.
			
			AZ, GD and PC contributed equally in the present work.
			
%

\pagebreak

\cleardoublepage

\widetext
\begin{center}
	\textbf{\large Supplementary Material for: 			Kibble-Zurek  {mechanism in driven-dissipative systems crossing a non-equilibrium phase}}
\end{center}

\vspace{8mm}

\setcounter{equation}{0}
\setcounter{figure}{0}
\setcounter{table}{0}
\setcounter{page}{1}
\renewcommand{\theequation}{S\arabic{equation}}
\renewcommand{\thefigure}{S\arabic{figure}}
\renewcommand{\bibnumfmt}[1]{[S#1]}
\renewcommand{\citenumfont}[1]{S#1}

\twocolumngrid

\

{In this Supplementary Information we provide a detailed account of the numerical and technical methods adopted in the main study, which are of crucial importance for the validation of our conclusions.}

\

\section{Pumping schemes and dynamical equation of the polariton field.}

{In order to validate the universality of the critical polariton dynamics described in the main text, we  perform two independent calculations for the two most celebrated pumping schemes for exciton-polaritons, {which} are schematically shown in \red{Fig.~\ref{fig:pump_scheme}}.  
	These correspond to the} optical parametric oscillation (OPO)  {scheme, whereby} polaritons are directly injected into the lower polariton band by the incident laser and then scatter into a coherent signal and idler population~\cite{savvidis2000angle,carusotto2013quantum}, and  {the incoherent pumping (IP) scheme, whereby} the high energy excitations generated by the incident light eventually relax and condense into the lower polariton band after a complex scattering sequence~\cite{kasprzak2006bose,deng2010exciton}.

\begin{figure}
	\includegraphics[width=\linewidth]{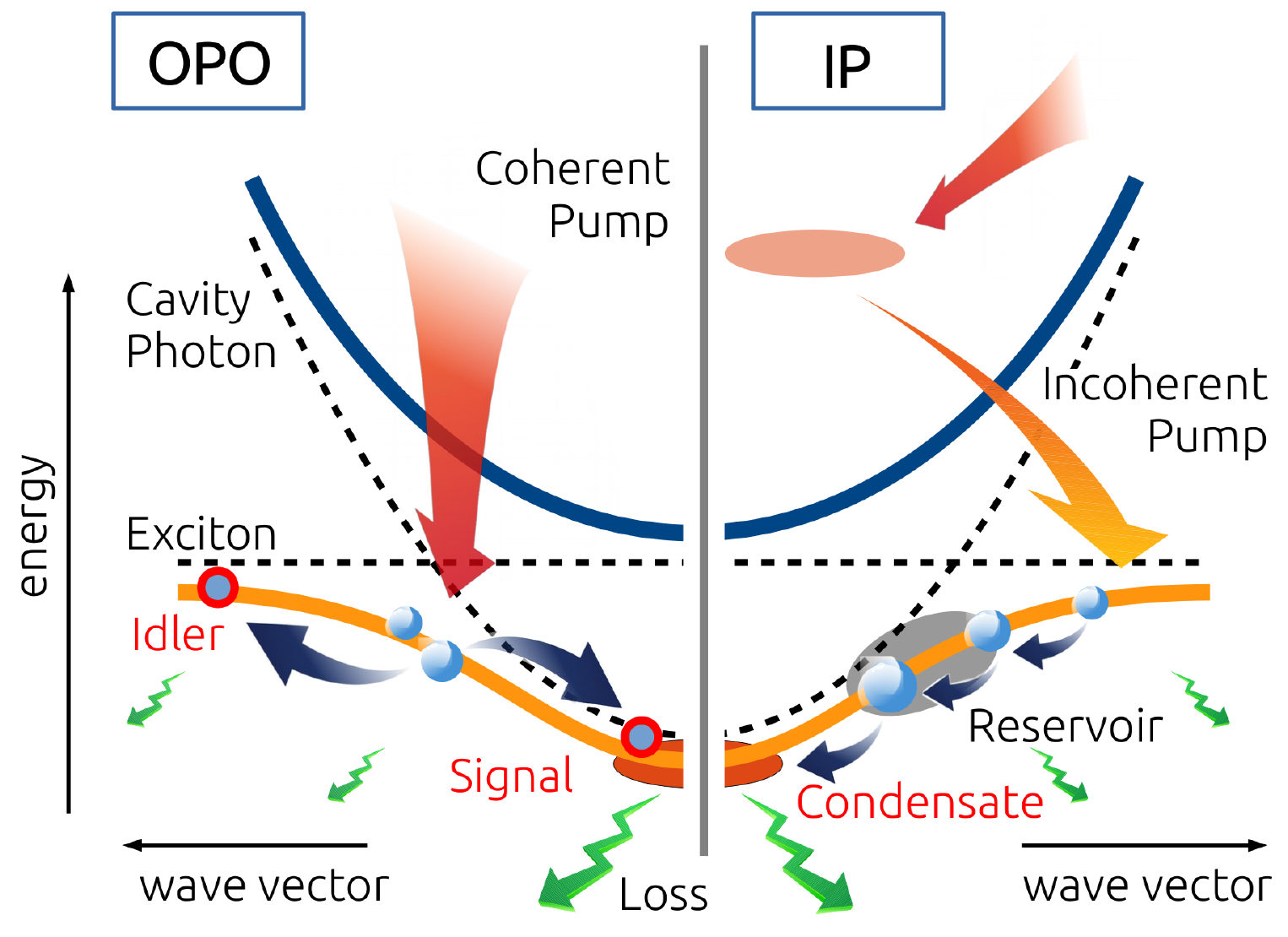}
	\caption{
		Schematic representation of the  two pumping schemes considered in the present study.
		In the OPO case (left picture), the external pump  {resonantly injects polaritons around the inflection point of the lower band} (red left arrow).  Resonant polariton-polariton scattering eventually generates two additional macroscopic occupied polariton modes: the signal (considered in the present study) and the idler.
		For the IP case (right picture), the external pump laser injects  high energy excitations such as electron-hole pairs in the semiconductor material. After some complex relaxation cascade process, these eventually produce a macroscopic density of polaritons at the bottom of the lower band.
	}
	\label{fig:pump_scheme}
\end{figure}

We describe the collective dynamics of the polariton fluid through a generalized stochastic Gross-Pitaevskii equation  {for the 2d polariton field as a function of the position $\bm{r}=(x,y)$ and time $t$.}
Equations of this kind can be derived by i) considering a truncated approximation of the evolution of the Wigner function  {of the polariton field}~\cite{carusotto2005spontaneous,PhysRevB.79.165302} or, alternatively,  {by} ii) treating the system within a Keldysh field path integral representation ~\cite{sieberer2016keldysh,PhysRevB.93.195306} and considering the Martin-Siggia-Rose formalism, which gives the Langevin equation for the system~\cite{keeling2017superfluidity,altland2010condensed}. 

Specifically, for the coherently pumped polaritons in the OPO regime, such an equation describes the dynamics of the system in terms of the exciton (X) and cavity photon (C) fields $\psi_\text{X,C}(\bm{r},t)$ (with $\hbar=1$)~\cite{dagvadorj2015nonequilibrium}:
\begin{equation}
i d \begin{pmatrix} \psi_\text{X} \\ \psi_\text{C} \end{pmatrix} =
dt\left[H_{\mathrm{MF}} \begin{pmatrix} \psi_\text{X} \\ \psi_\text{C} \end{pmatrix}
+ \begin{pmatrix} 0 \\ F_\text{p} \end{pmatrix}\right] +
\begin{pmatrix} \sqrt{\kappa_\text{X}} \: dW_\text{X} \\ \sqrt{\kappa_\text{C}} \:
dW_\text{C} \end{pmatrix}\;,
\label{eq:wigner}
\end{equation}
where  { $F_\text{p}= f_p e^{i (\bm{k}_\text{p} \cdot \bm{r}
		- \omega_\text{p} t)}$} denotes  {the coherent pump, which directly injects polaritons} at frequency $\omega_\text{p}$ and momentum $\bm{k}_\text{p}$, $\kappa_\text{X}$ and $\kappa_\text{C}$ are the decay rates of the excitons and photons respectively, and the thermal and quantum fluctuations are encoded in the complex-valued zero-mean white Wiener noise terms $dW_\text{A}$, which fulfill $\langle
dW^{*}_\text{l,A} (\bm{r},t) dW_\text{m,A} (\bm{r}',t) \rangle
=\delta_{\bm{r},\bm{r}'} \delta_\text{l,m}dt$, for $A=X,C$. 
The operator $H_{\mathrm{MF}}$ describes the dynamics at the mean-field level and takes the form:
\begin{displaymath}
H_{\mathrm{MF}}=
\begin{pmatrix} \frac{-\nabla^2}{2m_\text{X}} + g_\text{X}
(|\psi_\text{X}|^2-\frac{1}{dV}) - i\kappa_\text{X} & \frac{\Omega_\text{R}}{2} \\
\frac{\Omega_\text{R}}{2} & \frac{-\nabla^2}{2m_\text{C}} -i \kappa_\text{C}\end{pmatrix}
\; .
\end{displaymath} 
We neglect the kinetic term for the excitons $-\nabla^2/2m_\text{X}$, since $m_\text{X} \gg m_\text{C}$, where $m_\text{X}$ and $m_\text{C}$ are the exciton and cavity-photon masses respectively.
The Rabi-splitting $\Omega_R$ measures the  {radiative coupling} between excitons and photons; the  {coefficient $g_\text{X}$ of the nonlinear term quantifies the} exciton-exciton interaction strength;  {finally,} $dV=a^2$ is the element of volume of our 2D-grid, with lattice spacing $a$. 
As in the case of incoherently pumped polaritons, we consider a set of typical values for the parameters which can also be found in a large number of experimental setups {\cite{sanvitto2010persistent,dagvadorj2015nonequilibrium,carusotto2013quantum}}: $m_\text{C}=2.3\times10^{-5} m_e$
$\Omega_R\approx4.4 \mathrm{meV}$,  $g_X\approx 2 \times 10^{-3} \ \mathrm{meV\mu m^2}$. We consider $\kappa_\text{X}=\kappa_\text{C}$, with $\kappa_\text{C} = 1/6.58 \mathrm{ps}$. The external pump  {has a} momentum $\bf{k}_\text{p}=(1.6,0) \mathrm{\mu m^{-1}}$.  {Its frequency $\omega_p$ is chosen to be on resonance with the lower polariton band}, i.e.  $\omega_\text{p}=\omega(\bf{k}_\text{p})$.
We focus the study of the KZ mechanism on the dynamics of the signal polariton mode, which is obtained from the full polariton field from the numerical simulation of \eqref{eq:wigner} after a filtering process in momentum space around the signal mode~\cite{dagvadorj2015nonequilibrium,comaron2018dynamical}.

For the IP scenario, the equation describes the  {effective} dynamics of the lower polariton field $\psi=\psi(\bm{r},t)$  {and includes the complex relaxation process in a phenomenological way.} It reads as ($\hbar=1$)~\cite{PhysRevLett.99.140402,chiocchetta2013non,comaron2018dynamical}:
\begin{multline}
	\hspace{-4mm}id \psi = dt\bigg[ - \frac{
		\nabla^2}{2 m } +
	g|{\psi}|^2_{-} + \frac{i }{2}
	\bigg( \frac{P}{1+\frac{|{\psi}|^2_{-}}{n_\text{s}}} -
	\gamma \bigg) \\  +\frac{1}{2}\frac{P}{\Omega}\frac{\partial}{\partial t} \bigg]
	\psi +  dW
	\label{SGPE_IP}
\end{multline}
where $m$ is the polariton mass, $P$ is the strength of the homogeneous external drive, $g$ is the polariton-polariton interaction strength. The renormalized density $|{\psi}|^2_{-} \equiv
\left(\left|{\psi} \right|^2 - {1}/{2dV} \right)$  {includes} the subtraction of the Wigner commutator contribution (where $dV=a^2$ is the element of volume of our 2D grid with lattice spacing $a$), and $n_\text{s}$ is the saturation density. The zero-mean white Wiener noise $dW$ fulfils $\langle
dW^{*}(\vec{r},t) dW (\vec{r}',t) \rangle
=[(P+\gamma)/2] \delta_{\vec{r},\vec{r}'}dt$, where $\gamma$ is the inverse of the polariton lifetime.
{A frequency-selective pumping mechanism {has been implemented here in order to favour relaxation to low-energy modes~\cite{chiocchetta2013non,woutersLiew2010,wouters2010,comaron2018dynamical}}.}
In the present study, we use typical experimental parameters {\cite{nitsche2014algebraic}}: lifetime $\tau = 1/\gamma= 4.5 \mathrm{ps}$, $m = 6.2 \ 10^{-5} \ m_\text{e}$, $g = 6.82 \ 10^{-3} \ \mathrm{meV \mu m^2} $, {$\Omega = 11.09 \mathrm{ps^{-1}}$} and $n_\text{s}= 1500 \mathrm{\mu m^{-2}}$. 

A detailed explanation of the computational procedure adopted for the numerical integration of Eqs.~(\ref{eq:wigner})--(\ref{SGPE_IP}) is given below.

\

\section{Details of the numerical simulations. -}

We simulate the dynamics of the polariton system by numerically integrating  {in time} the stochastic differential equations for the polariton field  {shown in} \eqref{eq:wigner} for the OPO case and  {in} \eqref{SGPE_IP} for the IP case.
The numerical integration is performed on a 2d-lattice with periodic boundary conditions. In the OPO case, the 2D-grid is composed of $256^2$ lattice points, with lattice spacing $a=0.87\mu m$
and system size $L_\text{x}=L_\text{y}=222.72 \mathrm{\mu m}$. 
For the IP polariton system, the lattice is composed of $301^2$ grid points, with total lengths $L_\text{x} =L_\text{y}= 295.11 \mu m$ and lattice spacing $a=0.98\mu m$. 
Notice that the lattice spacing $a$,  {both} i) introduces a cut-off $\propto a^{-1}$ in the momentum representation of the field, and ii)  {is chosen} between a lower bound given by the macroscopic scale of the system, such as the healing length, and the upper bound $a^2 \gg g/\gamma$ given by the validity of the truncated Wigner methods used for the description of the stochastic field equations ~\cite{carusotto2013quantum,comaron2018dynamical,dagvadorj2015nonequilibrium}.

If not stated otherwise, methods and parameters of our simulations coincide with those in~\cite{comaron2018dynamical}: the time dynamics of the polariton field is performed by integrating \eqref{eq:wigner} and~\eqref{SGPE_IP}  {in time} with the XMDS2 software framework~\cite{dennis2013xmds2}. Specifically, we have used a fixed time-step of $0.3 \mathrm{ps}$ ($4.5\cdot 10^{-2} \mathrm{ps}$) for the OPO (IP) cases, which ensures stochastic noise consistency, and a fourth-order Runge-Kutta algorithm. 
All the results expressed in the present study are converged with respect to the number of stochastic realisations $N_{\textrm{stoch}}$. Specifically we consider $N_{\textrm{stoch}}=100$ (400) realisation for the case of the OPO (IP) polariton system.

\

%
\begin{figure*}[h]
	\includegraphics[width=0.6\linewidth]{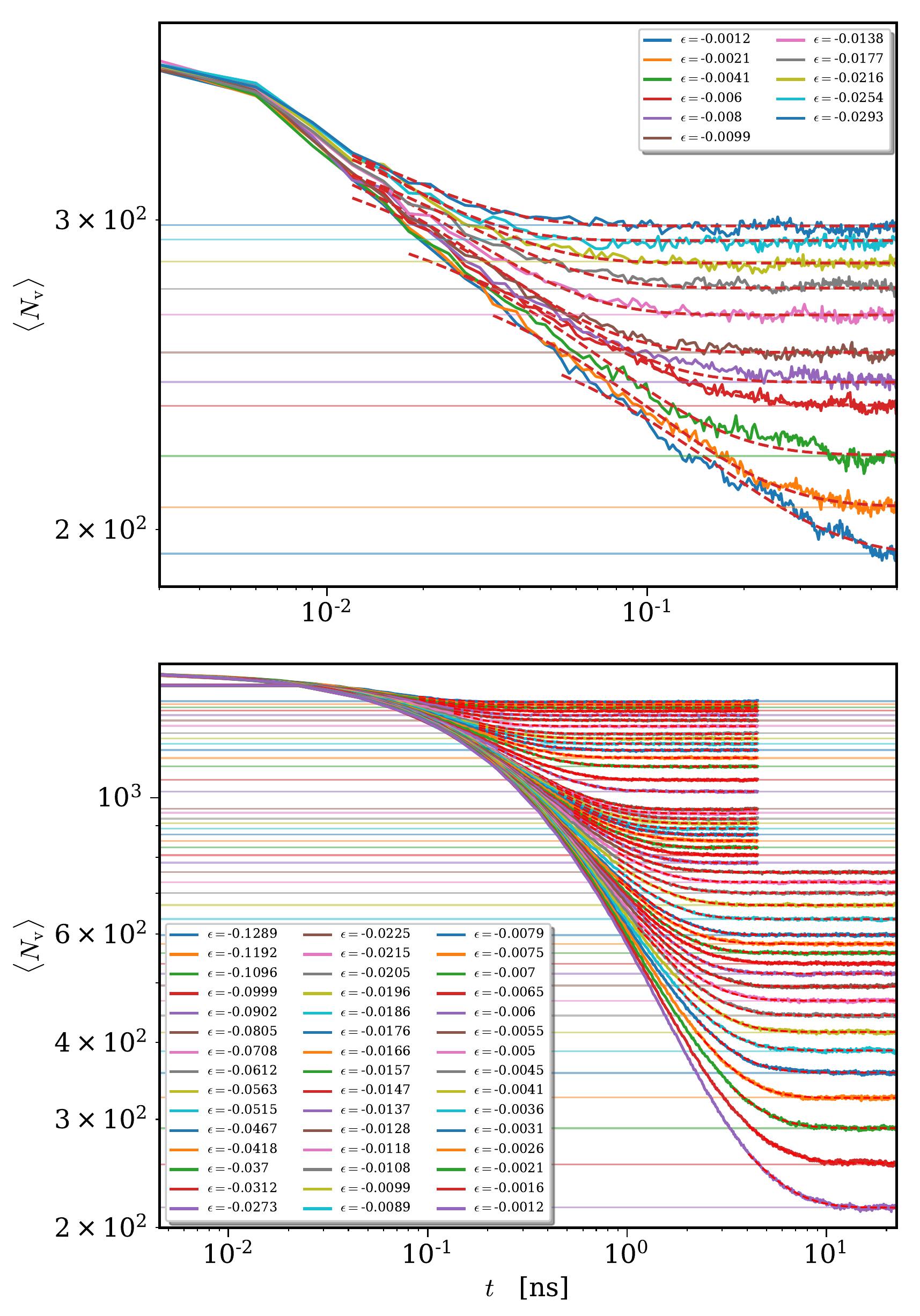}	
	\caption{\textbf{Numerical extraction of the relaxation timescale $\tau$.}
		Number of vortices as a function of time for different sudden rapid quenches into the disordered phase, both for the OPO (top panel) and the IP case (bottom panel). The horizontal lines indicate the steady state value $N_\mathrm{v}^\text{ss}(\epsilon)$ for each different value of $\epsilon$ calculated as described in the main text. 
		The red dashed curves indicate the fitting to the expression for the vortex number given by Eq.~\eqref{eq:tau_char_time}.
	}
	\label{fig:vortex_decay_tau}
\end{figure*}
%
\section{Finite-speed ramp protocol.}
{We conclude by discussing in detail the ramp protocol} considered in this work, which is given by \red{Eq.~1} of the main text. We  {bring} the polariton system from a disordered to a quasi-ordered phase across the critical value of the external  {pump intensity. The ramp speed is finite and inversely proportional to the characteristic ramp time $\tau_Q$.} 
The initial disordered phase is characterized by a  {vanishing} density of polaritons and  {a corresponding extremely high density of vortices in the stochastic field}. For the OPO, we study the upper threshold in the pump power and the initial disordered state is given by $\epsilon_i=-0.1653$, while for the IP case the initial distance to criticality takes the value $\epsilon_i=-0.1$.
{The critical values of the external pump read $f_{p}^{c} = 13.05 \mathrm{meV \mu m^{-1}}$ for the OPO case and $P_c=1.0325 \gamma$ for the IP case. The initial {and} final values of the pump intensity for the linear ramp 
	protocol have been chosen such that the constant $a_p$ appearing in \red{Eq.~1} of the main text is chosen to have comparable values
	$a_p \approx 0.2$. Specifically, our simulations are based on 
	i) {$f_{p}^{i} = 15.21 \mathrm{meV \mu m^{-1}} $} and {$f_{p}^{f} = 12.67  \mathrm{meV \mu m^{-1}}$} for the OPO and {ii) $P_i=0.93 \gamma$ and $P_f=1.14 \gamma$ for the IP}. Therefore, the constant $a_p$ takes the value 0.1942 (0.2) for the OPO (IP) case.}

The linear  {ramp} is completely determined by its  {characteristic timescale} $\tau_\text{Q}$
since the duration of the  {ramp} is $t_f-t_i=\tau_Q$ and the initial time is ${t_i}= \epsilon_\textrm{i} \tau_\textrm{Q} / a_p $, with $t_i<0$. 
\red{Fig.~2} of the main text displays the behaviour of the number of vortices as a function of time for different  {ramp speeds}.
We observe that the  {vortex number} monotonically decreases in time, as expected since we are  {bringing} the system from a  {highly} disordered phase to a  {well (quasi-)ordered phase.} 
The system crosses the critical point at $t=0$ and reaches the final stage of the evolution at ${t_f}= \epsilon_f \tau_\textrm{Q} / a_p $, which is a positive number. Note that $\epsilon_f=0.02894$ $(0.1)$ for the OPO (IP) case, where $\epsilon_f=\epsilon(t_f)$.

\

\section{Numerical extraction of the characteristic relaxation time for the vortices.}
\label{sec:etract_tau}
%
\begin{figure}
	\includegraphics[width=1.0\linewidth]{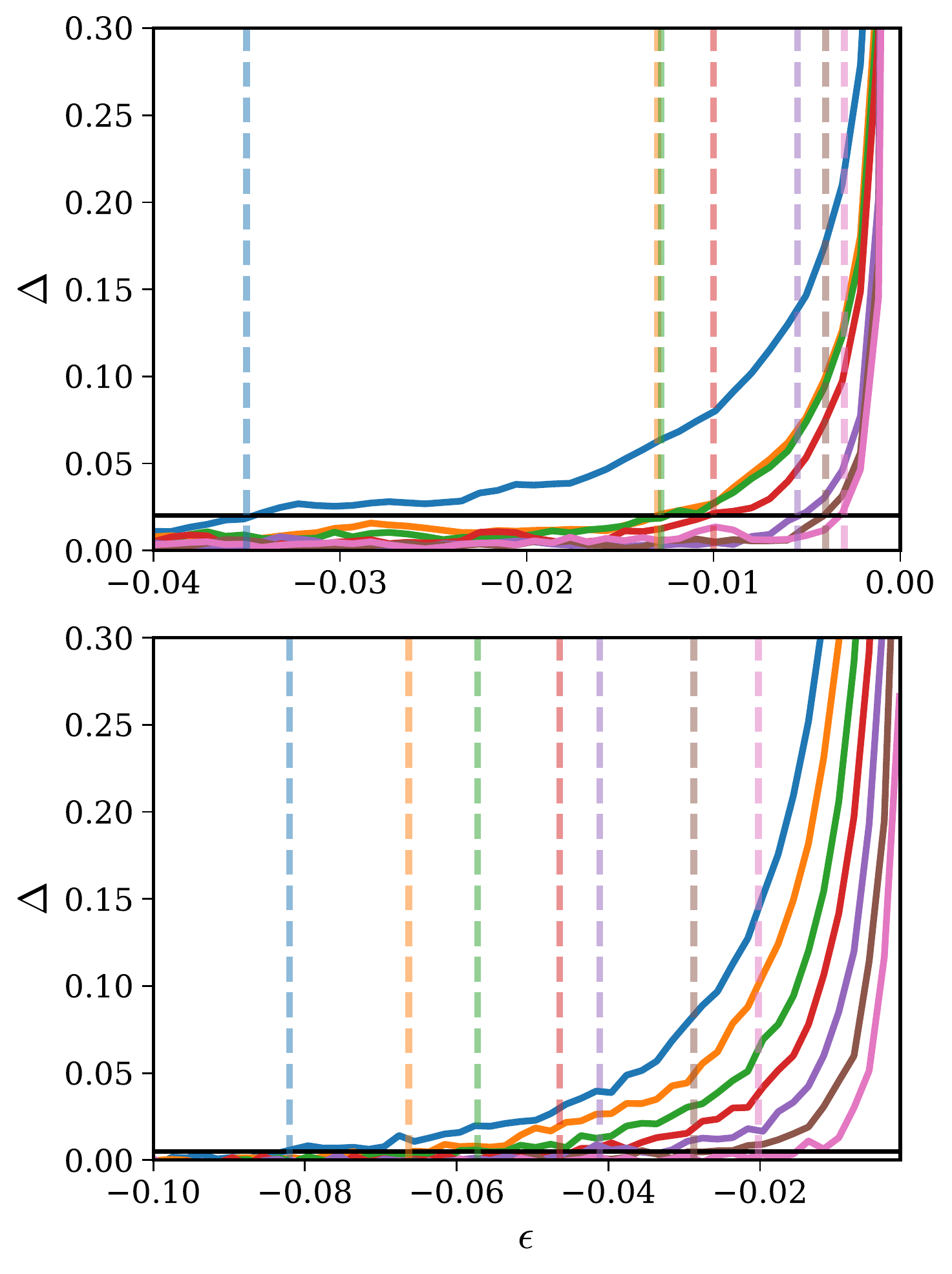}
	\centering
	\caption{\textbf{The difference between the number of vortices during the finite quench dynamics and at the steady-state.}
		$\Delta$ parameter for different  {ramp speeds}  (solid curves) as a function of $\epsilon$ for the OPO  (top panel) and for the IP system (bottom panel). The dashed vertical lines show the different crossover times  $\hat{\epsilon}_\text{num}(\tau_\text{Q})$, which satisfies $\Delta =\delta_\mathrm{v}$ with $\delta_\mathrm{v}$ ($\delta=0.02$  and $\delta=0.005$ for the OPO and IP case respectively). 
	}
	\label{fig:diff}
\end{figure}
%
\begin{figure}[h]
	\includegraphics[width=1.0\linewidth]{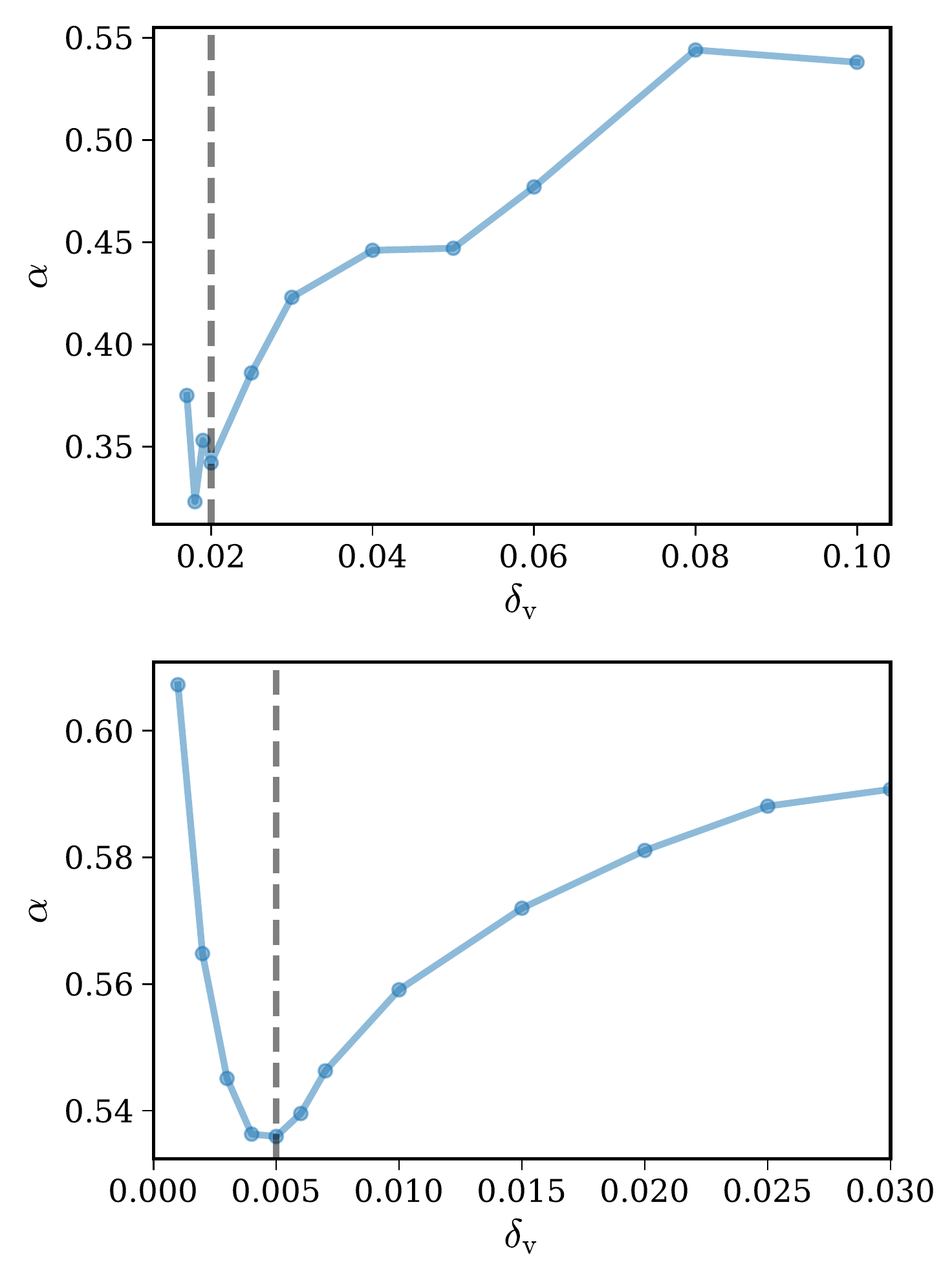}
	\centering
	\caption{\textbf{Obtaining the threshold $\delta_\mathrm{v}$ parameter.}
		To determine the best value for $\delta_\mathrm{v}$, we explore the behaviour of the exponent $\alpha$ -- extracted from the power-law fit of $\hat{t}_\mathrm{num}(\tau_\mathrm{Q})$ (see Fig.~\ref{fig:t_num_vs_tauQ}) -- as a function of the parameter $\delta_\mathrm{v}$. 
		We display the exponents obtained in both OPO (upper panel) and IP regime (bottom panel). 
		For OPO we consider $\delta_\mathrm{v}=0.02$ in our study since we can observe that the value of the exponent $\alpha$ converges as this specifica value of $\delta_\mathrm{v}$.
		For IP case, the exponent $\alpha$ decreases for $\delta_\mathrm{v} > 0.005$, before a divergence when $\delta_\mathrm{v}\to 0$. Thus, the value $\delta_\mathrm{v} = 0.005$, corresponding to the minimum of the curve, is chosen as the best choice for the analysis described in main text.
	}
	\label{fig:deltav}
\end{figure}

%
\begin{figure*}[t!]
	\centering
	\includegraphics[width=.8\linewidth]{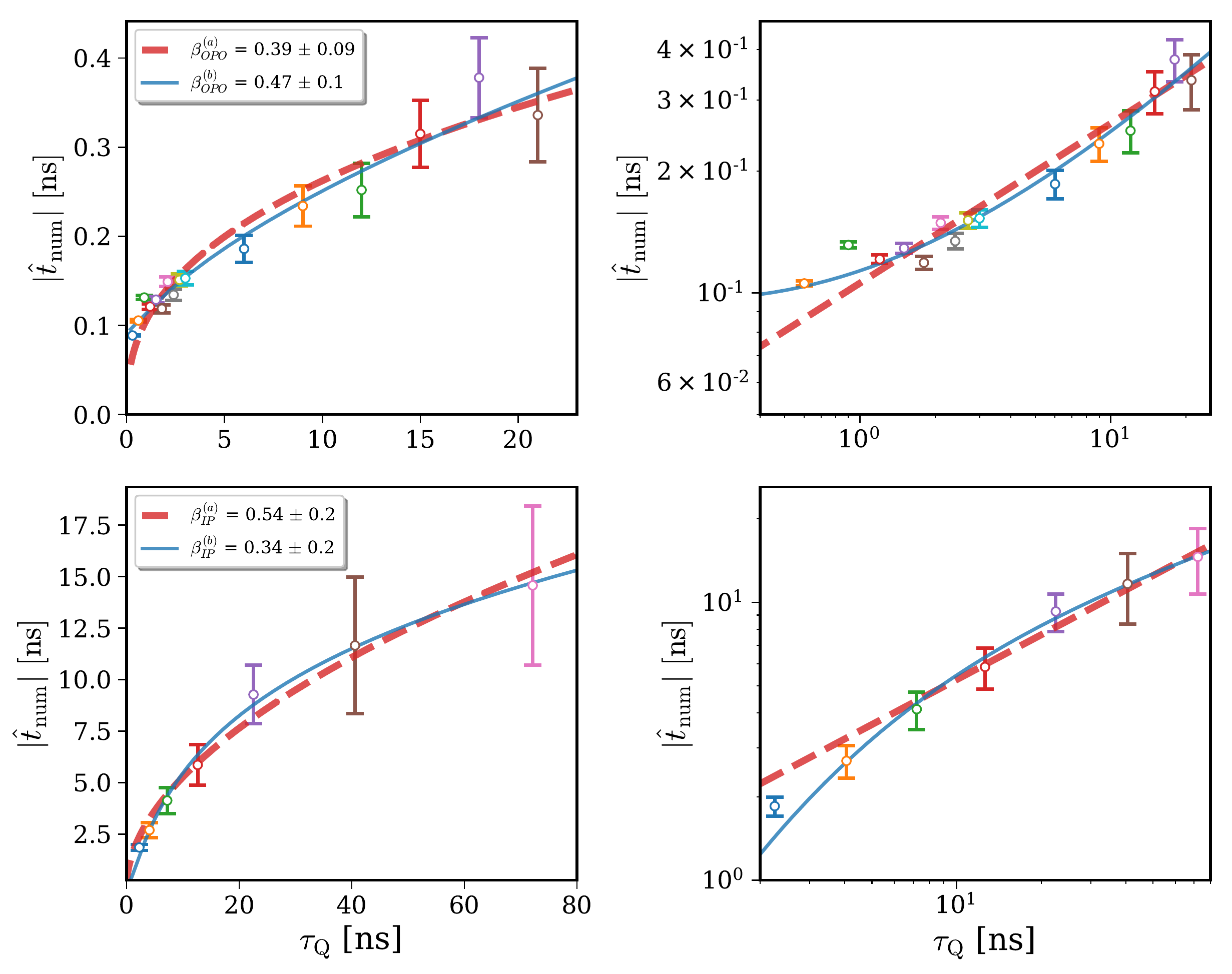}
	\caption{\textbf{``Numerical'' crossover time.}
		An absolute value of $\hat{t}_\text{num}$ as a function of the quench rate, both for the OPO system (top panels) and the IP system (bottom panels) {, plotted in linear-linear (left panels) and log-log (right panels) scales}. We observe that, as expected,  {$|\hat{t}_\text{num}|$} is a monotonically increasing function with respect to the quench rate $\tau_Q$. 
		 {Lines are fits according to the two fitting strategies discussed in Sec.~\ref{sec:scaling_crossover}, namely a power law (red dashed) and a BKT-like law (blue solid).}
		{Error bars are extracted taking into account the uncertainty arising from the calculation of the intersection between $\Delta$ and $\epsilon(\tau_Q)$ depicted in Fig.~\ref{fig:diff}}.
	}
	\label{fig:t_num_vs_tauQ}
\end{figure*}
\begin{figure*}[t!]
	\centering
	\includegraphics[width=.8\linewidth]{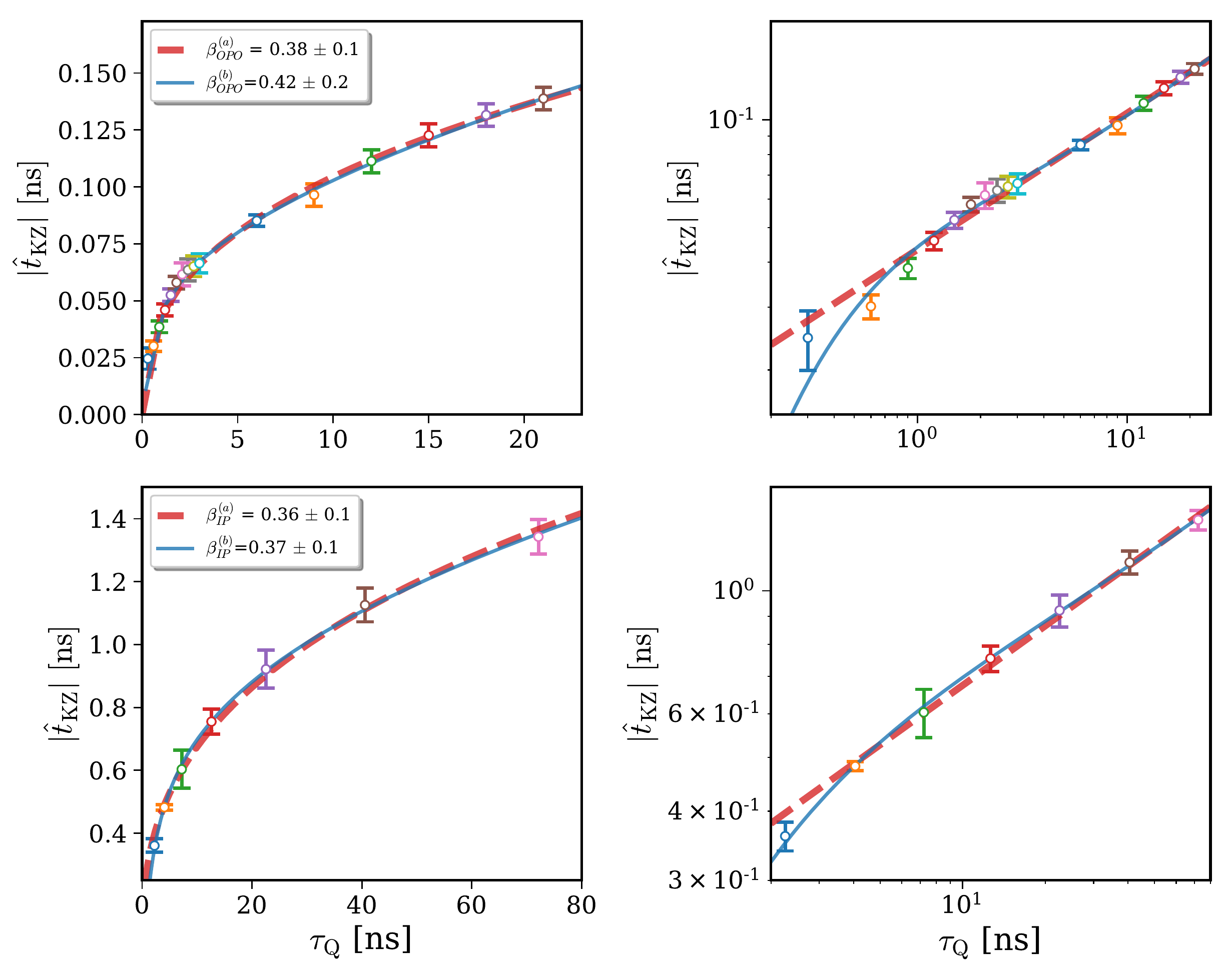}
	\caption{\textbf{``Predicted'' crossover time.}
		An absolute value of $\hat{t}_\textrm{KZ}$ as a function of the quench rate $\tau_Q$, both for the OPO (top panels) and the IP (bottom panels) system {, plotted in linear-linear (left panels) and log-log (right panels) scales}. As expected, the crossover time is a monotonically increasing function of the quench rate.
		 {Lines are fits according to the two fitting strategies discussed in Sec.~\ref{sec:scaling_crossover}, namely a power law (red dashed) and a BKT-like law (blue solid).}
		{Error bars are extracted considering the uncertainty arising from the calculation of the intersection between the relaxation time $\tau$ and time $t$ as shown in Fig.~\ref{fig:tau_vs_time}.}
	}
	\label{fig:t_pred_vs_tauQ}
\end{figure*}
%
%
In this section we discuss in detail how we obtain the characteristic time $\tau(\epsilon)$ of the relaxation of the vortices towards the steady state  {for the specific value of $\epsilon$ we are interested in. Note that $\tau(\epsilon)$ is only well defined within the disordered phase, while it diverges $\tau\to\infty$ in the} quasi-ordered phase~\cite{comaron2018dynamical}.
Firstly, we need to establish the initial configuration from which we are quenching the system towards the  {chosen} value $\epsilon$. 
{For the OPO case, this initial state coincides with the initial states for the finite quench protocol described in the main text i.e. $\epsilon_i=-0.1653$, while $\epsilon_i=-1$ for the IP case. }
Note, that this initial state is located  {farther away in the disordered phase}, i.e. $|\epsilon|<|\epsilon_i|$.
Then we \emph{suddenly} quench the system from $\epsilon_i$ to $\epsilon$ and let it evolve until it reaches the steady state at  {the final $\epsilon$} (see Fig.~\ref{fig:vortex_decay_tau}).

In order to obtain the characteristic relaxation time $\tau(\epsilon)$ for the vortices, we follow the following two steps:
i) We numerically estimate the average number of defects in the steady state at a given $\epsilon$.
This is obtained by probing the average difference between each temporal point $N_\mathrm{v}(t_\textrm{j})$ and their corresponding values of adjacent time steps, $N_\mathrm{v}(t_\textrm{j-1})$ and $N_\mathrm{v}(t_\textrm{j+1})$, where $t_{j}$ is the time a the $j-$th step of the temporal evolution. 
When this difference is lower than the fluctuation strength $\sigma = \sqrt{(N_\textrm{p})}$, the average density of points $N_\mathrm{v}({t})$ is considered to be at steady state.
This allow us to obtain the curve $N_\mathrm{v}^\textrm{ss}(\epsilon)$ depicted in {\red{Figs.~1 and 2}} in the main text as red thick line.
ii) We assume that the vortex dynamics after the sudden quench in the disordered region follows an exponential-type of relaxation towards the steady-state values at the end of the evolution:
\begin{equation}
N_\mathrm{v}(t,\epsilon) \sim N_\mathrm{v}^\text{ss}(\epsilon) + a_{\tau} e^{-\frac{t}{\tau(\epsilon)}},
\label{eq:tau_char_time}
\end{equation}
where the characteristic vortex time $\tau(\epsilon)$ coincides with the characteristic time of the exponential relaxation and $a_\tau$ is a free parameter. As a result, we obtain the characteristic time $\tau(\epsilon)$ displayed in \red{Fig.~3} of the main text.

\

\section{Numerical extraction of the ``numerical'' crossover time.}
\label{sec:comp_pred}
Inspired by previous work~\cite{{jelic2011quench}}, in order to determine the
\emph{numerical} crossover time, we look for the point at which the number of vortices $N_\mathrm{v}$ at each finite quench departs from the steady-state vortex number $N_\mathrm{v}^\text{ss}$.
Therefore, as shown in \red{Fig.~2} of the main text, we need to compare $N_\mathrm{v}(\epsilon,\tau_\text{Q})$ and $N_\mathrm{v}^\text{ss}(\epsilon)$ curves. From this comparison we  {obtain} the detaching point $\hat{\epsilon}_{\textrm{num}}$, which can be easily  {converted} to the crossover time $\hat{t}_\textrm{num}$ by considering expression (1) in the main text of the paper. Note that this relation also permits to describe the vortex number $N_\mathrm{v}$ during the finite quench dynamics as a function of $\epsilon$ instead of $t$, as it appears in \red{Fig.~2} of the main text.

In order to determine,  {for each $\tau_Q$,} the departure point $\hat{\epsilon}_\mathrm{num}$ (dashed vertical lines in \red{Fig.~2} of the main text), we introduce  a parameter $\Delta$ that quantifies the difference between $N_\mathrm{v}$ and $N_\mathrm{v}^\text{ss}$:
\begin{equation}
\Delta(\tau_\text{Q},\epsilon) = \frac{\left|N_\mathrm{v}(\tau_\text{Q},\epsilon) - N_\mathrm{v,fit}^\text{ss}(\epsilon)\right|}{N_\mathrm{v,fit}^\text{ss}(\epsilon)}.
\end{equation}
$N_\mathrm{v,fit}^\text{ss}$ is the  numerical fit to $N_\mathrm{v}^\text{ss}$ with $N_\mathrm{v,fit}^\text{ss}(\epsilon)\sim \exp( c \epsilon^{b})$ 
where $b$ and $c$ are  fitting parameters.

We compute the $\Delta$ parameter for each finite quench at all times ($\epsilon$) (see Fig.~\ref{fig:diff}). As expected, we observe that far from the critical region, i.e. at the beginning and at the intermediate stages of the time evolution, $\Delta \approx 0$. This is a clear indication that the system is in the adiabatic regime. However, there is a moment in the evolution where $\Delta$ starts to increase. This behaviour reflects the fact that the system leaves the adiabatic regime and enters into the impulse regime.

Therefore, the crossover point, i.e. either $\hat{\epsilon}_{\textrm{num}}$ or $\hat{t}_{\textrm{num}}$, is given by the point from which $\Delta \neq 0$. Since there is a certain ambiguity in establishing that point in the numerical solutions, we introduce a threshold $\delta_\mathrm{v} \ll 1$ such that the system is considered to lie in the adiabatic regime for $\Delta < \delta_\mathrm{v}$, and behave non-adiabatically for $\Delta > \delta_\mathrm{v}$.  {Specifically here we use} $\delta_\mathrm{v}=0.02$ and $\delta_\mathrm{v}=0.005$ for the OPO and IP polariton system respectively (see Fig.~\ref{fig:deltav}).
Consequently, the  {detaching} time $\hat{\epsilon}_\text{\textrm{num}}$ is thus taken at the intersection point $\Delta = \delta_\mathrm{v}$ (see Fig.~\ref{fig:diff}). The results of $\hat{t}_\textrm{num}$ as a function of $\tau_Q$ are shown in Fig.~\ref{fig:t_num_vs_tauQ}.
%
\begin{figure}
	\includegraphics[width=1.0\linewidth]{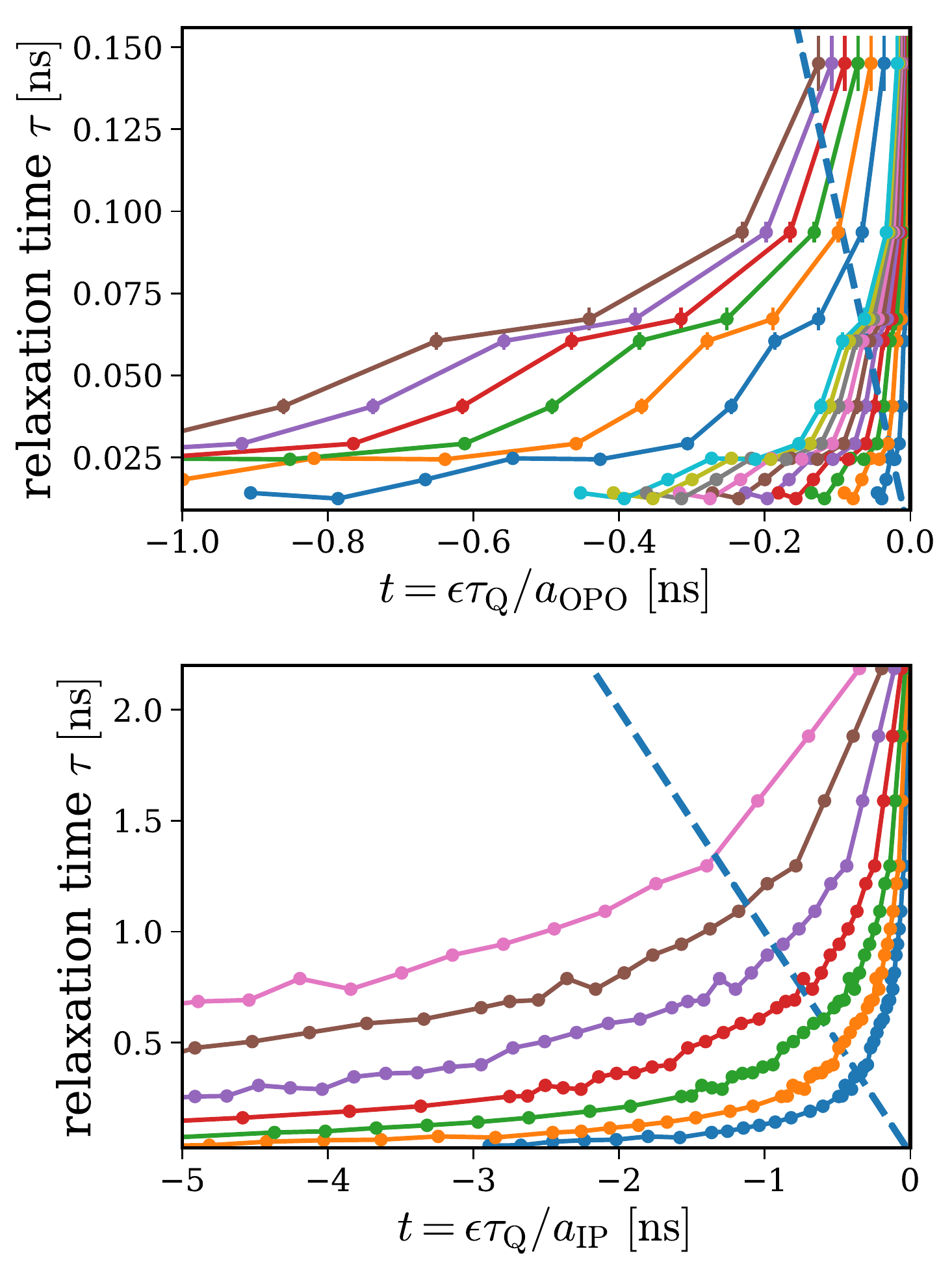}
	\caption
	{
		\textbf{An estimate of the crossover time from the characteristic relaxation time of the vortices.}
		Characteristic relaxation time $\tau$ for the vortices as a function of the quench time $t$ for different finite quenches (coloured curves) both for the OPO (top panel) and IP (bottom panel) pumping schemes. Dashed straight line indicates $\tau=|t|$.  The intersection between the coloured curve and the straight line gives the predicted crossover time $\hat{t}_{\textrm{KZ}}$.
	}
	\label{fig:tau_vs_time}
\end{figure}
\section{{Extraction} of the ``predicted'' crossover time  {from the Kibble-Zurek hypothesis}.}
\label{sec:theor_pred}
In this section we describe the procedure performed in order to numerically obtain the crossover point  $\hat{t}_\textrm{KZ}$, as predicted by the KZ mechanism.

Firstly, for all different finite quenches considered in the present work, we obtain an expression for the characteristic relaxation of the vortices $\tau$ as a function of the quench time $t$, i.e. $\tau=\tau(t)$. This is done by combining our numerical estimate of $\tau$ as a function of the criticality $\epsilon$ (see previous section and \red{Fig.~2} of the main text) with the expression of the finite quench given in \red{Eq.(1)} of the main text, which states $t = \epsilon \tau_\textrm{Q} / a $. Therefore we evaluate $\tau$ as a function of $t$ for each different finite quench, i.e. $\tau \left ( \epsilon(t) \right ) \to \tau(t)$, as shown in \red{Fig.~2} of the main text.

Secondly, we consider Zurek's relation i.e. \red{Eq.~(2)} of the main text, which is reduced in our case to $\tau(\epsilon(\hat{t}_\textrm{KZ}))=\hat{t}_\textrm{KZ}$. Consequently, we obtain the crossover time $\hat{t}_\textrm{KZ}$ for each different finite quench by determining the intersection point between curves $\tau(t)$ and the straight line $|t(t)|$, as shown in Fig.~\ref{fig:tau_vs_time}.
As expected, the slower the quench the more stretched the curve $\tau(t)$ is. This fact eventually results in higher values of $\hat{t}_\textrm{KZ}$, as expected for slower quenches falling out of the adiabatic regime earlier than faster ones. The dependence of $\hat{t}_\textrm{KZ}$ as a function of $\tau_Q$ is shown in Fig.~\ref{fig:t_pred_vs_tauQ}.

\begin{figure*}
	\centering
	{\includegraphics[width=.49\linewidth]{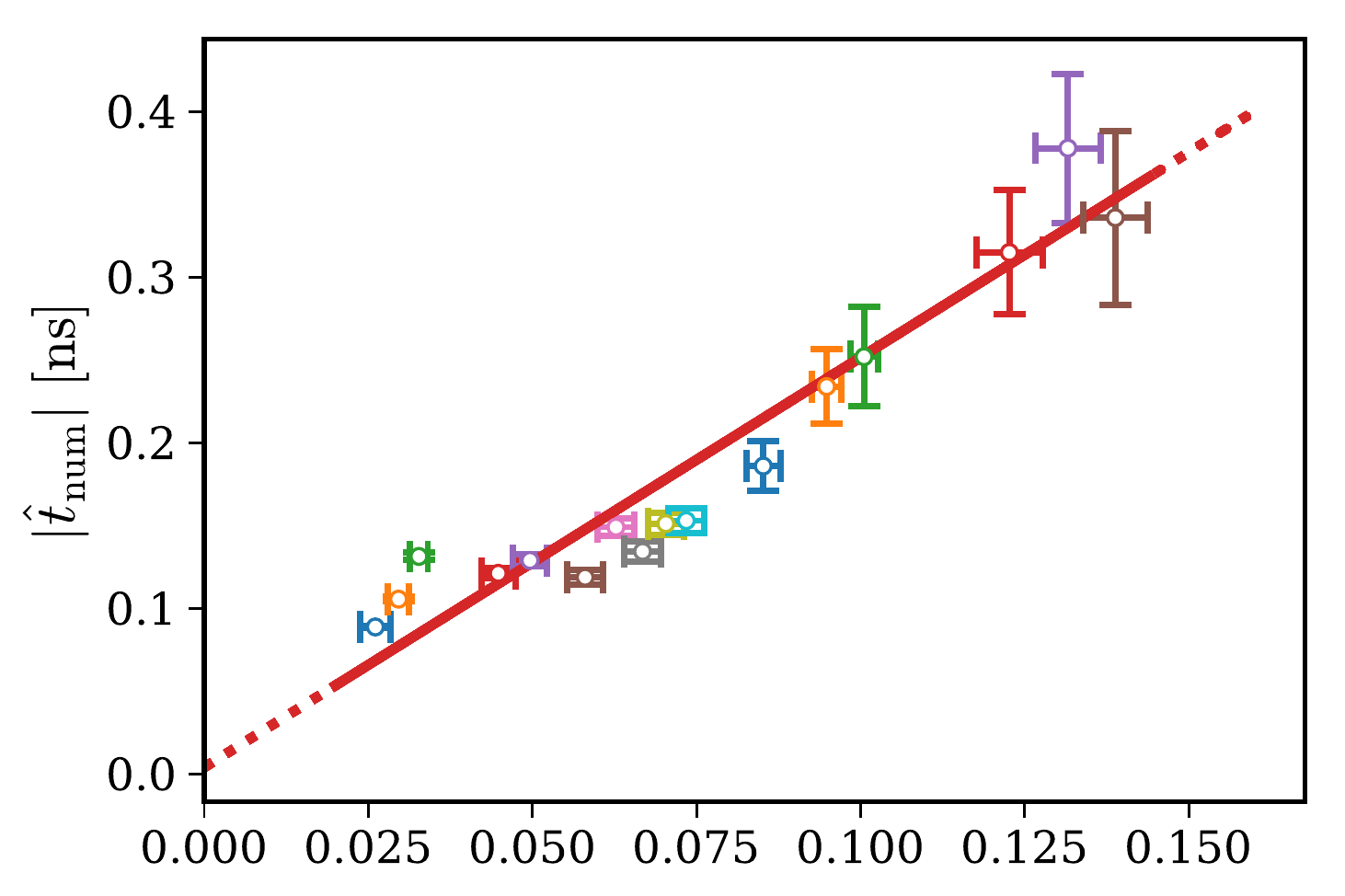}}
	{\includegraphics[width=.49\linewidth]{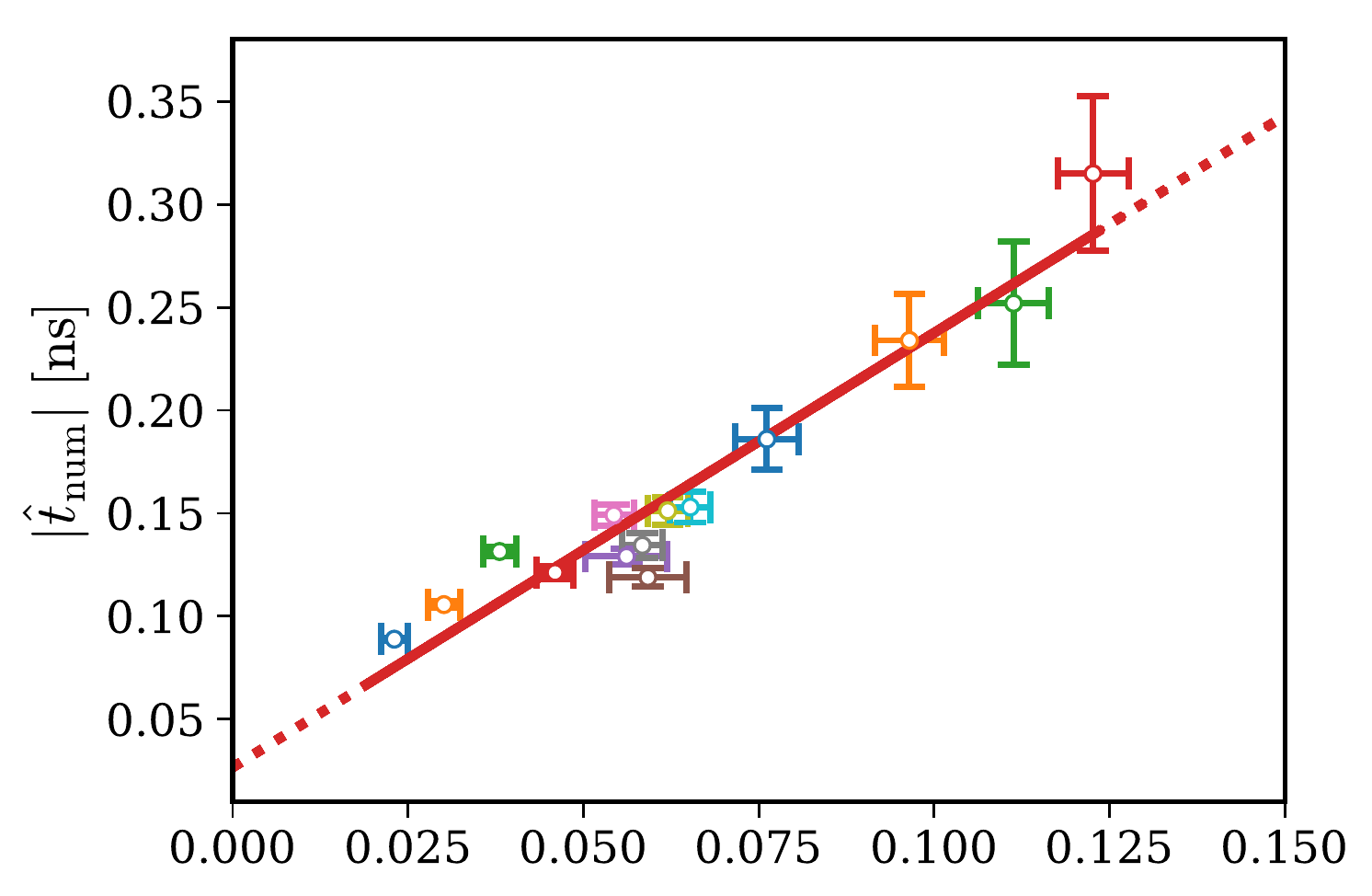}} {\includegraphics[width=.49\linewidth]{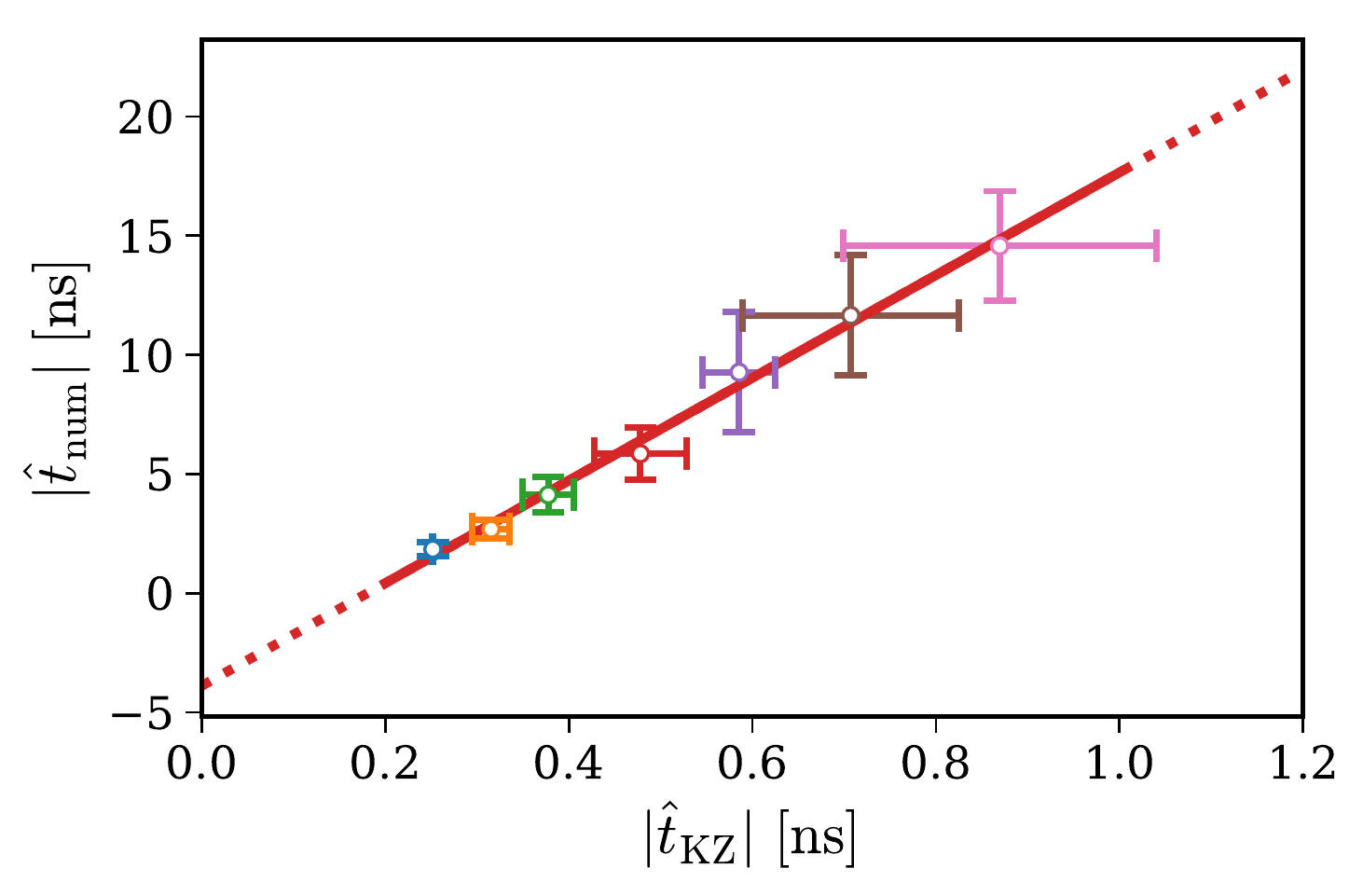}}
	{\includegraphics[width=.49\linewidth]{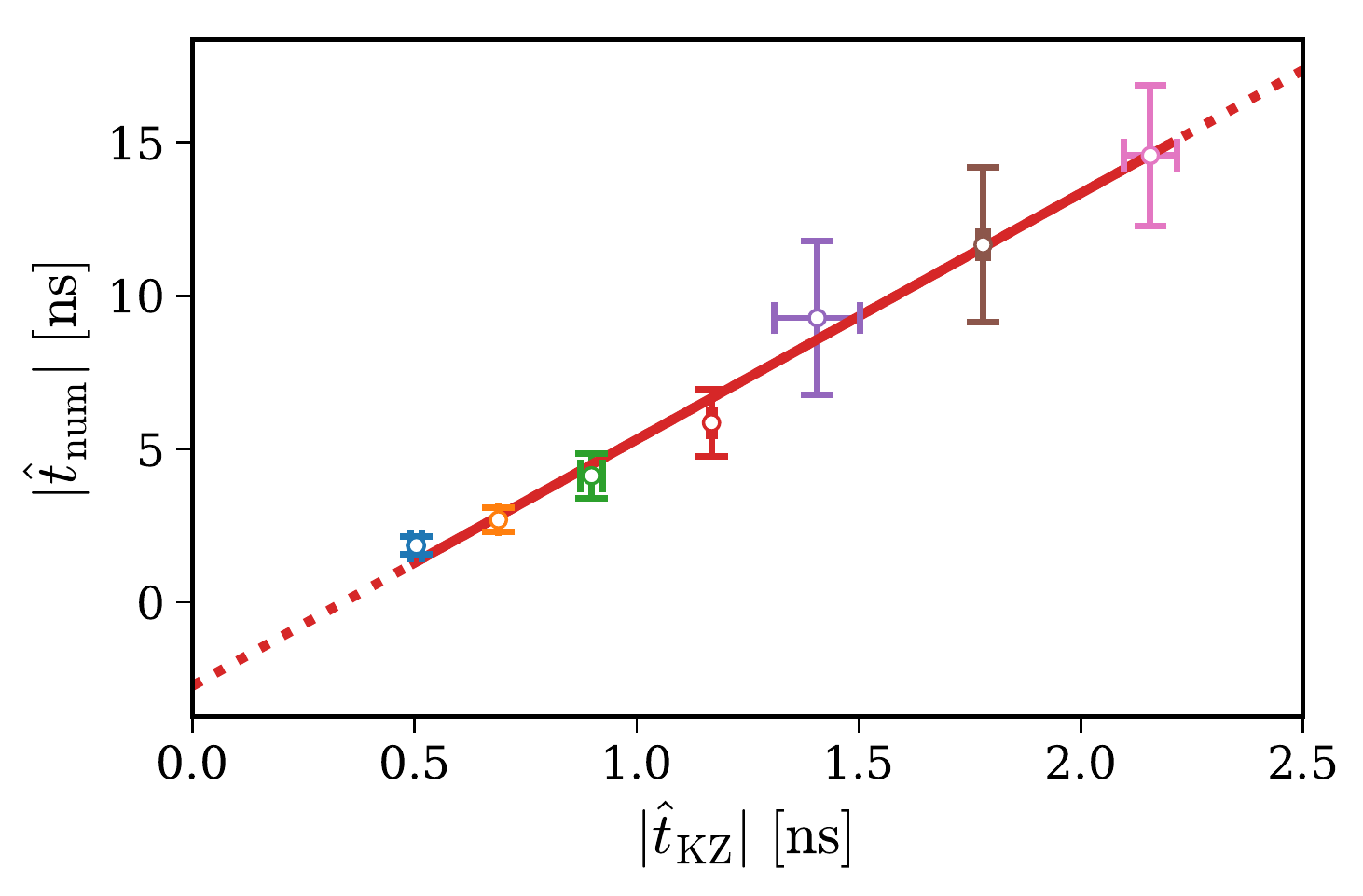}}
	\caption{\textbf{Independence of the KZ mechanism from the non-universal constant `A' for OPO (top) and IP (bottom).}
		Numerical crossover time $|\hat{t}_\mathrm{num}|$ as a function of the theoretical crossover time predicted by Zurek's relation $|\hat{t}_\mathrm{KZ}|$ expressed in \red{Eq.~(2)} of the main paper. $A=0.6$ (top left) and $A=1.6$ (top right) for the OPO.  $A=0.5$ (bottom left) and $A=2$ (bottom right) for the IP system in all sets of data.  {We observe a  linear relation between the numerical and theoretical crossover times, as predicted by Zurek's hypothesis} (\red{Eq.(2)} in the main paper). Such linear relation holds also for other different values of $A$  {in the probed range} $0.5\leq A \leq 2.0$.}
	\label{fig:different_A_OPO}
\end{figure*}
%

 {
	\section{The scaling relation between crossover time and quench rate.}
	\label{sec:scaling_crossover}
	In this section we briefly discuss the scaling relation between the crossover time $\hat{t}$ and quench rate $\tau_\mathrm{Q}$.
	The dependence of $\hat{t}_\textrm{num}$ ($\hat{t}_\textrm{KZ}$) on $\tau_Q$, presented in Fig.~\ref{fig:t_num_vs_tauQ} (Fig.~\ref{fig:t_pred_vs_tauQ}), shows the expected monotonic increase  with respect to the quench rate.
	We tentatively extract the scaling exponents by fitting the numerical data with two characteristic functions: $a)$ a simple 
	power-law 
	\begin{equation}
	\hat{t} \sim \tau_\mathrm{Q}^{\beta},
	\end{equation}
	and $b)$ with 
	\begin{equation}
	\hat{t} \sim \tau_\mathrm{Q}\left(\frac{z}{\log \left( \frac{\tau_\mathrm{Q}}{\tau_0} \left(   \frac{\ell_0 z }{ \log(\frac{\tau_\mathrm{Q}}{\tau_0})}  \right)^{(1+\nu)/\nu} \right) }\right)^{1/\nu},
	\end{equation}
	equivalent to Eq.~(24) of Ref~\cite{jelic2011quench}, i.e. an asymptotic prediction for the critical scaling for the BKT transition (therefore including logarithmic corrections), as discussed in  \cite{jelic2011quench,dziarmaga2014quench}. 
	Here $\ell_0\sim 1$ is an universal constant and $\tau_0$ some microscopic time scale \cite{jelic2011quench,comaron2018dynamical}.
	The fitting curves are plotted in Figs.~\ref{fig:t_num_vs_tauQ}~and~\ref{fig:t_pred_vs_tauQ} as red dashed and blue solid lines, respectively.
	From the outcome of the fits we find that the statistical sample of data points used in the present work is not sufficiently large neither to clearly identify the best fit, nor to obtain a quantitative prediction of the critical exponents, which is  beyond the scope of this work.
}

 {
	However, one can still attempt to extract ``rough'' estimates of the critical exponents.
	We henceforth label the exponent extracted by fit $(a)$ as $\beta^{(a)}$, and note that a corresponding critical parameter $\beta^{(b)}$ can be extracted by means of fit $(b)$ through the expression  $\beta = \nu z /(1+ \nu z)$, where $\nu$ is the static and $z$ the dynamic critical exponents \cite{hohenberg1977theory}. 
	Note, that the extracted $\beta$ values (reported in legends of Fig.~\ref{fig:t_num_vs_tauQ} and Fig.~\ref{fig:t_pred_vs_tauQ}) show a remarkable agreement when comparing the ``numerical" with the ``predicted" crossover time ($\hat{t}_{\mathrm{num}}$ and $\hat{t}_{\mathrm{KZ}}$) for both OPO and IP systems.
	Moreover, from fit $(b)$, we find an approximative values for the dynamical critical exponent $z=2.0(5)$, in very good agreement with our recent findings \cite{comaron2018dynamical}.
	Values of $\beta$ extracted from the two fitting curves, lie in the approximate window $0.34 < \beta < 0.54$, and are therefore
	broadly consistent with the predicted value for mean-field theory $\beta_\mathrm{MF}=0.5$ \cite{hohenberg1977theory}.
}
 {
	Concluding, we stress that no concrete conclusions on the values of the critical exponents or scaling behaviour should be drawn from the preliminary analysis discussed above.
	Potential corrections due to the non-equilibrium nature of the critical point, or associated with the asymptotic behaviour characterizing the scaling relation in the BKT framework \cite{jelic2011quench,dziarmaga2014quench}, could be present beyond the finite size and quench duration times accessible in the numerics presented in this work.
	In future, it will be interesting to further investigate the exact values of the critical exponents by means of more accurate finite-size scaling analysis.
}

\section{About the non-universal constant `A' appearing in Zurek's relation.}
\label{sec:about_A}
{We now show that} our results both in the OPO and IP systems are independent of the non-universal constant `A' appearing in the Zurek's relation (see \red{Eq.~(2)} in the main text), which accounts for the microscopic details.
We find that the linear relation between the numerical and the predicted `crossover time' holds for a vast range of different values of the `$A$' constant, particularly $0.5 \leq A \leq 2$ (note that the results presented in the main text are for  $A=1$).
For the OPO (IP), we find that the slope of such a linear relation ranges from 2.496 (21.54) to 1.811 (7.67) for the $A=0.5$ and $A=2.0$ cases respectively. 
We also find the average value of the intercept to be $50.23 \pm 89.66$ for the OPO, i.e. zero intercept lies within the error bars. 
For the IP case, we find the average value of the intercept to be $3.42 \pm 0.5$.

In Fig.~\ref{fig:different_A_OPO} we show  $A=0.6$ and $A=1.6$ cases for the OPO  (top left and right, respectively), 
and $A=0.5$ and $A=2$ for the IP system (bottom left and right, respectively).
For the OPO, we find that the linear fits get worse in both limits $A\to 0.5$ and $A\to 2$, since the lower limit excludes slow quenches (which are the ones that sustain KZ phenomenon), and the upper limit only accounts for very slow quenches, where finite size problems of the numerical simulations close to the critical point can arise. 
Similar behaviour is found in the limits of small and large `$A$' for the IP case.

\end{document}